\documentclass[aps,prb,twocolumn,amsmath,amssymb,floatfix,showkeys,showpacs,reprint]{revtex4-2}
\usepackage{graphicx}
\usepackage{bm} % bold math
\usepackage{color}
\usepackage{xcolor}
\usepackage{graphics} % <------- for LatTeX + PS
\usepackage{graphicx} % Include figure files
\usepackage{epsfig}
\usepackage{amssymb}
\usepackage{amsmath} %,amssymb}
\usepackage{amsthm}
\usepackage{amstext}
\usepackage{dcolumn} % Align table columns on decimal point
\usepackage{makeidx}
\usepackage{subfigure}
\usepackage{float}

\begin{document}
%\selectlanguage{english}

\title{Peculiarities of Rabi oscillations and free induction decay in two-component nuclear spin systems with Ising nuclei interactions}

\author{G.A. Rusetsky}

\affiliation{Scientific-Practical Material Research Centre, Belarus National Academy of Sciences, 19 P. Brovka str., Minsk 220072 Belarus}

\author{V.M. Kolesenko}

\affiliation{Scientific-Practical Material Research Centre, Belarus National Academy of Sciences, 19 P. Brovka str., Minsk 220072 Belarus}

\author{U. Shymanovich}
%\date{\today}

\begin{abstract}
Nuclear spin systems in manganites, having separation of ferromagnetic and antiferromagnetic phases, which are manifested as two lines in the NMR spectrum, are studied. Taking into account the Ising nuclear interaction, we obtain an analytical description of Rabi oscillations and free induction decay in two-component nuclear spin systems when a radio-frequency pulse non-resonantly excites the nuclei of one magnetic phase and resonantly the nuclei of the other phase. It is revealed that the nonlinearity of interaction between the nuclei of these phases results in additional harmonics in the Rabi oscillations of the non-resonantly excited subsystem. We also show that the Ising interaction inside this subsystem forms multiple echoes in the free induction decay, whereas their non-monotonic damping and amplitude modulation is caused by the spin coupling between the subsystems.
\end{abstract}

%\keywords{Manganites, Ising interaction, spin-spin interaction, two-component system, Rabi oscillations, free induction, multiple single-pulse echo signals}

\maketitle
\section{INTRODUCTION}
A number of papers \cite{Zangara.2015,Zangara.2017,Kaur.2013,Sanchez.2014,Guerry.2017,Morgan.2012} is devoted to studying the dynamics of coherent NMR signals in solids which are generated when a nuclear spin system is exposed to a one or two radio frequency (RF) pulses. Special attention is paid there to the nuclear spin-spin interactions. This interest is due to the perspective of using such spin systems for quantum information processing and quantum simulation \cite{Alvarez.2015,Bernien.2017,Dominguez.2016,Mizushima.2017}.

If one considers a nuclear system of magnetically ordered materials then it is necessary to take into account the spin-spin interactions which occur via the electronic subsystem (Suhl-Nakamura interaction) \cite{BorovikRomanov.1984}. An example of such a spin system is manganite. They have very intriguing electronic and magnetic properties (colossal magnetoresistance, magnetocaloric effect etc.). Therefore manganites are studied quite a lot and they are very promising for practical use in the future \cite{Papavassiliou.2000,Panopoulos.2018,Germov.2019}. Strong competition between the double exchange, superexchange interactions of $\mathrm{Mn^{4+}}$ and $\mathrm{Mn^{3+}}$ ions and electron-phonon interactions (or Jahn-Teller effect in Mn) in manganites can lead to the appearance of antiferromagnetic, ferromagnetic insulating, and metallic phases. For example, by means of the two-pulse (Hahn) echo one was able to register (I) the $\mathrm{{}^{55}Mn}$ NMR lines from localized $\mathrm{Mn^{4+}}$, $\mathrm{Mn^{3+}}$ and $\mathrm{Mn^{2+}}$ ions which correspond to the ferromagnetic insulating phase and (II) the $\mathrm{Mn^{4+}}$, $\mathrm{Mn^{3+}}$ line associated with the ferromagnetic metallic phase \cite{Tomka.1998,Mazur.2012,Savosta.2001}.

Single-pulse echo (SPE) can also be generated in manganites \cite{Zviadadze.2013}. The SPE in magnetically ordered materials differs from that in materials without such ordering where, in particular, two SPE signals registered in toluene were used to determine the chemical shift of protons of the benzene and methyl groups \cite{Kuzmin.2001,Kaiser.1981}. 
Firstly, the SPE is strong in magnetic materials with a domain structure. This is due to the amplification effect caused by the hyperfine interaction of the electronic and nuclear subsystems \cite{BorovikRomanov.1984,Kuzmin.2012,Kuzmin.2005}.
Secondly, the SPE formation in magnetically ordered materials is significantly influenced by the nuclei spin-spin interaction which takes place via the spin ordered electron system. This indirect nuclear-nuclear interaction (Suhl-Nakamura) \cite{Pincus.1968,Tagirov.2014} is stronger than the usual dipole-dipole interaction between nuclear magnetic moments and leads to a large dynamic shift of the NMR frequency during the exposure to the RF pulse. In this case (I) the NMR frequency depends on the amplitude of the oscillations in the nuclear spin system and (II) the nuclear spin echo is formed due to the frequency modulation mechanism \cite{BorovikRomanov.1984}.
This mechanism was used to explain multiple single-pulse echoes (MSPE) \cite{Bunkov.1974}. These echoes were experimentally observed in an easy-plane antiferromagnetic $\mathrm{MnCO_3}$ at time moments $2\tau$ and $3\tau$ after the primary single-pulse echo (PSPE) appearing at the moment $\tau$ (the time is counted from the end of the RF pulse, $\tau$  is the pulse duration). Similar MSPE signals were also observed and theoretically described as a part of the free induction decay (FID) after exposure semimetals, ferrites, and manganites to one and two RF pulses \cite{Zviadadze.2013,Zviadadze.2015,Akhalkatsi.2002}. However, the nature of these signals remains not fully understood because there is no analytical description of the shape and oscillation's phase for each echo in the MSPE \cite{BorovikRomanov.1984,Akhalkatsi.2002,Mamniashvili.2015,Shakhmuratova.1997}.

The SPE signals in two-component nuclear systems were theoretically studied neglecting the spin-spin interaction of nuclei \cite{Kaiser.1981,Kuzmin.2001,Kuzmin.2006}. However, for example in manganites with two magnetic phases (ferromagnetic and antiferromagnetic), this approximation is not applicable. Furthermore, features of FID in nuclear systems are determined by the dynamics of the Rabi oscillations (transient nutation)\cite{Fedoruk.2002,Saiko.2018,Saiko.2018b,Khasanov.2007,Khasanov.2003}, which can have their own peculiarities in manganites. Moreover, the spin-spin interaction of nuclei in manganites within one magnetic phase and between the phases has different effects on the SPE signal formation.

Therefore, in order to describe MSPE signals in manganite it is necessary to take into account the strong spin-spin interaction of nuclei within the magnetic phases and between them. For this purpose, we investigate the Rabi oscillations and the FID in the two-component nuclear system with the Ising interaction. It will allow us to better understand the mechanisms of the MSPE formation and obtain important spectroscopic information about systems with spin-spin interaction.
\section{Theory}
Let us consider a system that consists of two nuclei, each having a spin of $1/2$. We suppose these nuclei are placed in different static magnetic fields $B_{n1} $ and $B_{n2}$ which are parallel to the axis $z$. Let us apply an RF field having amplitude ${B_R}$ and frequency $\omega$ to these nuclei in the direction perpendicular to the $z$ axis. In the general case, the spin-spin interaction between nuclei can be described in the framework of the XYZ Heisenberg model. However, to simplify calculations we use an Ising model \cite{Schmitt.2016, Schmitt.2018, ernst1990principles}, which takes into account the interaction of only the $z$ components of the spins and modulates the energy states of the spin system. The interaction of the $x$ and $y$ components of the spins causes modulation of the amplitude of the exciting field. We assume that frequency modulation has the greatest influence on the dynamics of the spin system, and we neglect amplitude modulation. The Hamiltonian of such a system can be written as:
\begin{equation} \label{eq1}
\hat{H}={{\hat{H}}_{1}}+{{\hat{H}}_{2}}+{{\hat{H}}_{I}},
\end{equation}
where 
\begin{equation*}
{{\hat{H}}_{1,2}}={{\omega }_{n1,n2}}{{\hat{S}}^{z}_{1,2}}+\frac{1}{2}{{{\eta }_{1,2}}{{\omega }_{R}}}({{\hat{S}}^{+}_{1,2}}{{e}^{-i\omega t}}+{{\hat{S}^{-}}_{1,2}}{{e}^{i\omega t}}) 
\end{equation*}
are Hamiltonians, describing interaction of the nuclei with the RF field,${{\hat{H}}_{I}}=\tilde{k}{{\hat{S}}^{z}_{1}}\cdot {{\hat{S}}^{z}_{2}}$ is a Hamiltonian of the Ising nuclear interaction, ${{\hat{S}}^{\pm, z}_{1,2}}$ are the spin operator components, ${{\omega }_{n1,n2}}=\gamma {{B}_{n1,n2}}$ are the  Larmor precession frequencies of nuclei, ${{\omega }_{R}}=\gamma {{B}_{R}}$ is the Rabi frequency, $\gamma$ is gyromagnetic ratio, ${{\eta }_{1,2}}$ are the gain factors of RF field on nuclei, $\tilde{k}$ is the Ising interaction coefficient (between nuclei).

\begin{figure}[htp]
\begin{tabular}{cc}
\includegraphics[width=80mm]{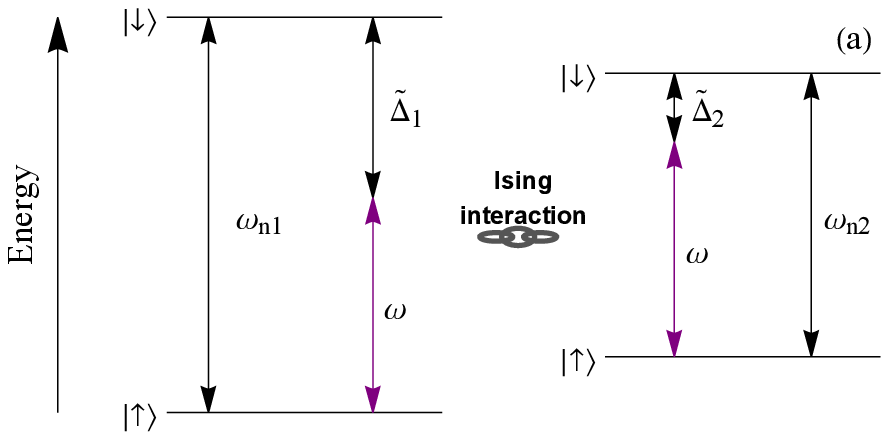}\\
\includegraphics[width=80mm]{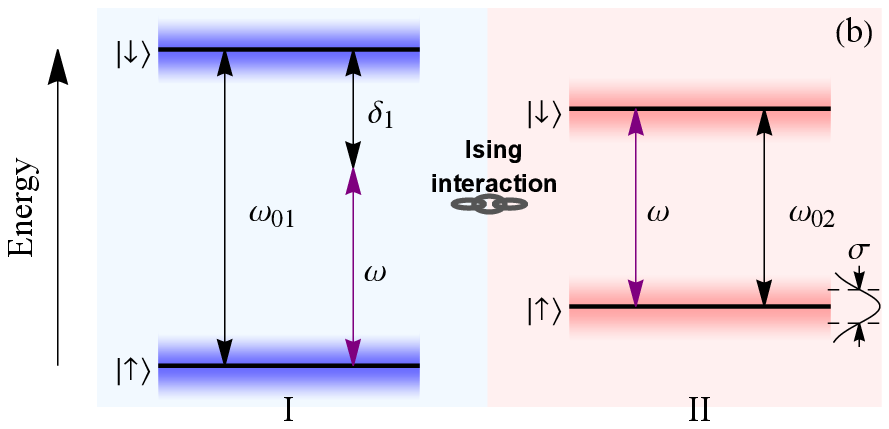}
\end{tabular} 
\caption{\label{fig:my0}Energy diagram of a pair of nuclei (a) and nuclei of two subsystems (b) in a magnetic field and coupled via the Ising interaction. The 2nd nuclear subsystem is excited by a resonant RF field with a frequency $\omega$. The nuclei of the pair have resonant frequencies $\omega_{n1}$ and $\omega_{n2}$. The detunings of these nuclei from the carrier frequency of the RF field are $\tilde{\Delta}_1$ and $\tilde{\Delta}_2$. The frequencies of the nuclei of the first and second subsystems are distributed according to the Gaussian function with HWHM $\sigma$ near the central frequencies of the spectroscopic transitions $\omega_{01}$ and $\omega_{02}$, respectively. The central frequency of the nuclei of the first subsystem differs from the frequency of the exciting field by $\delta_1$.}
\end{figure}

The evolution of the system under investigation can be described by the Heisenberg equation
\begin{equation} \label{eq2}
i\dot{\hat{o}}=[\hat{o},\hat{H}],
\end{equation}where $\hat {o}$ is an arbitrary operator which does not explicitly depend on time.
Let us transform \eqref {eq2} to the frame, rotating around the $z$ axis with frequency $\omega$: ${{\hat{o}}_{R}}= \exp (-i{{\hat{S}}^{z}_{2}}\omega t) \exp (-i{{\hat{S}}^{z}_{1}}\omega t)\hat{o}\exp (i{{\hat{S}}^{z}_{1}}\omega t) \exp (i{{\hat{S}}^{z}_{2}}\omega t)$. Using \eqref {eq1} and \eqref {eq2} we can get
\begin{equation} \label{eq3}
{{\dot{\hat{o}}}_{R}} = i\omega [{{\hat{o}}_{R}},{{\hat{S}}^{z}_{1}}] + i\omega [{{\hat{o}}_{R}},{{\hat{S}}^{z}_{2}}]-i[{{\hat{o}}_{R}},{{\hat{H}}_{R}}],
\end{equation}
where
\begin{equation} \label{eq0}
{{\hat{H}}_{R}}={{\omega }_{n1}}\hat{S}_{1}^{z}+{{\omega }_{n2}}\hat{S}_{2}^{z}+{{\eta }_{1}}{{\omega }_{R}}\hat{S}_{1}^{x}+{{\eta }_{2}}{{\omega}_{R}}\hat{S}_{2}^{x}+\tilde{k}\hat{S}_{1}^{z}\cdot \hat{S}_{2}^{z}
\end{equation}
and ${{\hat{o}}_{R}}$ are the Hamiltonian and the arbitrary operator in the rotating frame, correspondingly. In the Fig. \ref{fig:my0}(a) the energy diagram of a pair of nuclei under consideration is shown.
Using Eq. \eqref {eq3} and taking into account the commutation relations we obtain the equations of motion for the corresponding spin operator components

%\begin{subequations}
%\label{eq4}
\begin{eqnarray}
\label{eq4}
{{\dot{\hat{S}}^{x}_{1,2}}} & = &-{{\tilde \Delta }_{1,2}}{{\hat{S}}^{y}_{1,2}}-\tilde{k}{{\hat{S}}^{y}_{1,2}}\cdot {{\hat{S}}^{z}_{2,1}},
   \nonumber\\
   {{\dot{\hat{S}}^{y}_{1,2}}} & = &{{\tilde \Delta }_{1,2}}{{\hat{S}}^{x}_{1,2}}-{{\omega }_{R}}{{\eta }_{1,2}}{{\hat{S}}^{z}_{1,2}}+\tilde{k}{{\hat{S}}^{x}_{1,2}}\cdot {{\hat{S}}^{z}_{2,1}},
    \\
{{\dot{\hat{S}}^{z}_{1,2}}} & = & {{\omega }_{R}}{{\eta}_{1,2}}{{\hat{S}}^{y}_{1,2}},\nonumber
\end{eqnarray}
%\end{subequations}
where ${{\tilde \Delta }_{1,2}}={{\omega }_{n1,n2}}-\omega$.

If we replace the Eqs. \eqref{eq4} for operators with the corresponding equations for their mean values, then we get in the first approximation
%\begin{subequations}
%\label{eq5}
\begin{eqnarray}
\label{eq5}
\langle {\dot{\hat{S}}^{x}_{1,2}}\rangle &=& -\langle {{ \hat S^{y}_{1,2}}}\rangle ({{\tilde \Delta }_{1,2}}+\tilde{k}\langle {{ \hat S^{z}_{2,1}}}\rangle ),\nonumber
   \\
\langle {\dot{\hat {S}}^{y}_{1,2}}\rangle & = &-{{ \eta }_{1,2}}{{\omega }_{R}}\langle {{ \hat S^{z}_{1,2}}}\rangle +\langle {{ \hat S^{x}_{1,2}}}\rangle ({{\tilde \Delta }_{1,2}}+\tilde{k}\langle {{ \hat S^{z}_{2,1}}}\rangle )	,
   \\
 \langle {\dot{\hat{S}}^{z}_{1,2}}\rangle &=& {{\eta }_{1,2}}{{\omega }_{R}}\langle {{ \hat S^{y}_{1,2}}}\rangle.\nonumber
\end{eqnarray}
%\end {subequations}

Using \eqref{eq5} we can find the following equations for the magnetic moment components ${{u }_{1,2}} =m_0\gamma \langle {{ \hat S}^{x}_{1,2}}\rangle$, ${{\upsilon }_{1,2}} =m_0 \gamma \langle {{ \hat S}^{y}_{1,2}}\rangle$, ${{w}_{1,2}} = m_0\gamma \langle {{ \hat S}^{z}_{1,2}}\rangle$:
%\begin{subequations}
%\label{eq6}
\begin{eqnarray}
\label{eq6}
{{\dot{u}}_{1,2}}&=&-{{\upsilon }_{1,2}}({{\tilde \Delta }_{1,2}}+k{{w}_{2,1}}),\nonumber
   \\
{{\dot{\upsilon }}_{1,2}}&=&-{{\omega }_{R}}{{ \eta }_{1,2}}{{w}_{1,2}}+{{u}_{1,2}}({{\tilde \Delta }_{1,2}}+k{{w}_{2,1}}),
   \\
{{\dot{w}}_{1,2}}&=&{{\omega }_{R}}{{\eta }_{1,2}}{{\upsilon }_{1,2}},\nonumber
\end{eqnarray}
%\end {subequations}
where $k=\tilde{k}/(m_0 \gamma)$, $m_0$ is the magnetization in equilibrium (without RF field).

\begin{figure}[htp]
\begin{tabular}{cc}

\includegraphics[width=40mm]{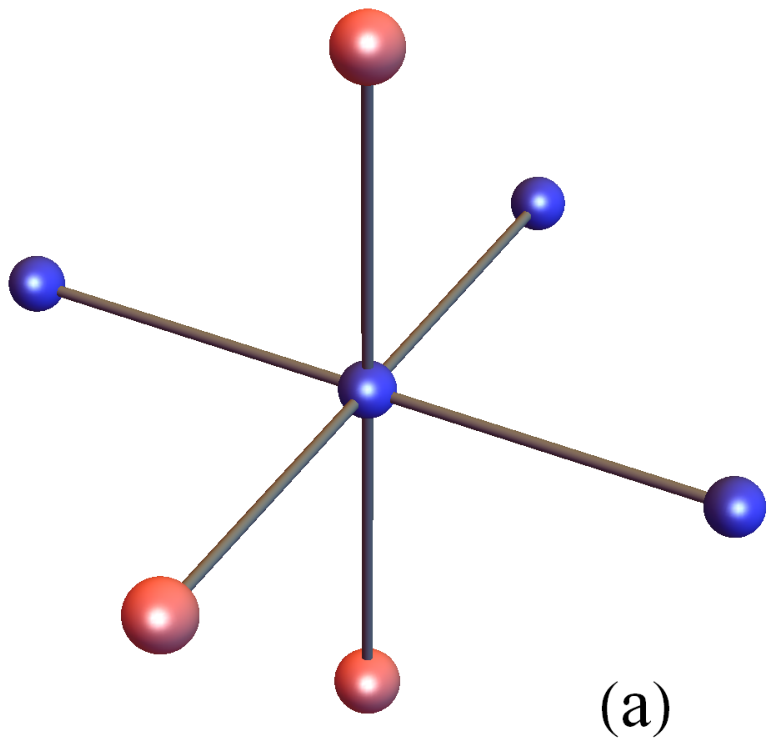}
&
\includegraphics[width=40mm]{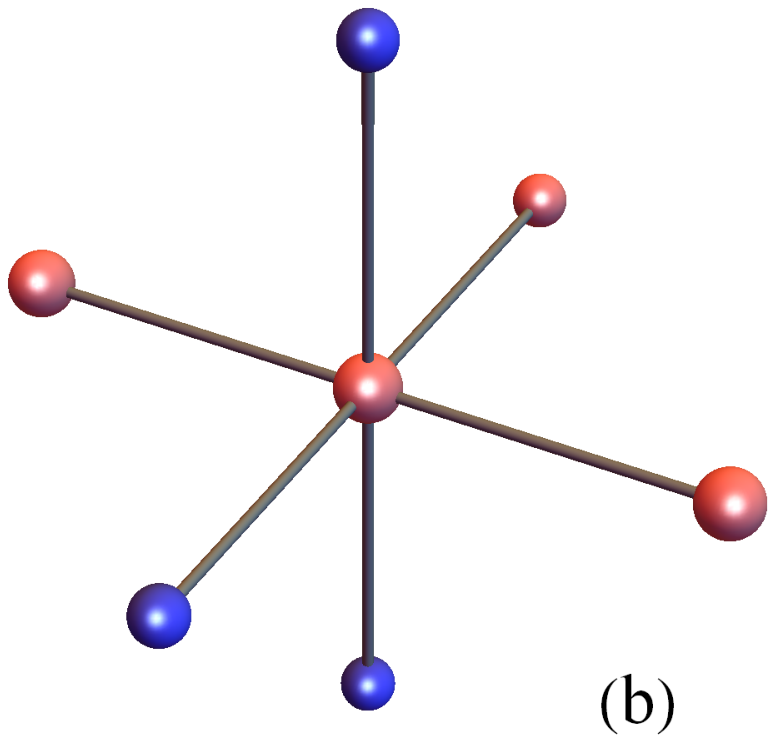}\\
\includegraphics[width=40mm]{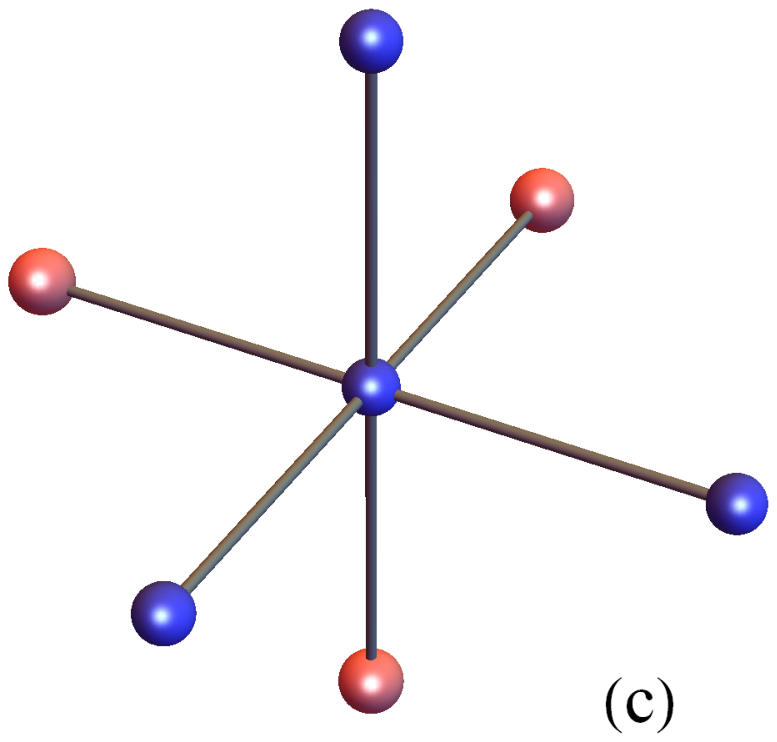}
&
\includegraphics[width=40mm]{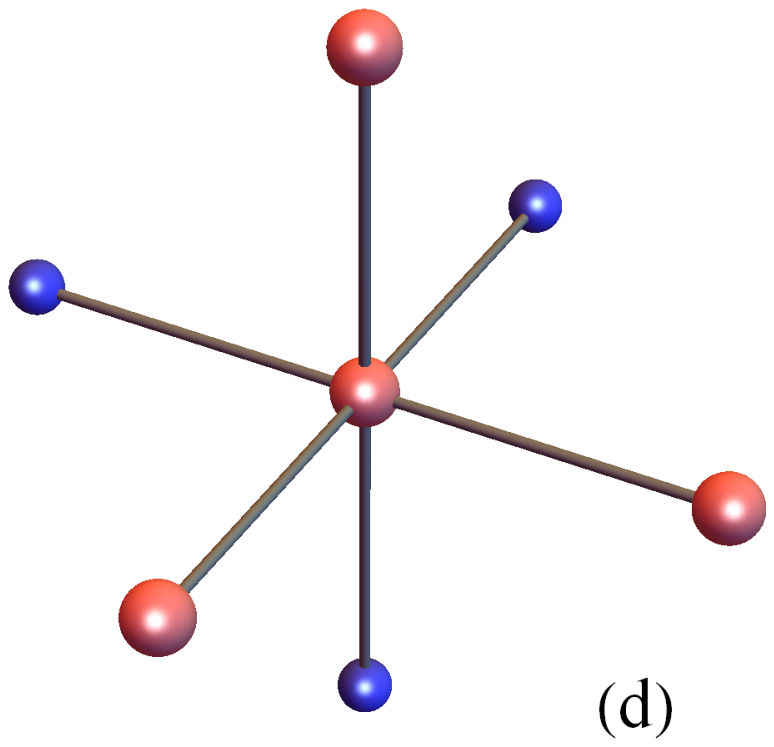}\\
\multicolumn{2}{c}{\includegraphics[width=60mm]{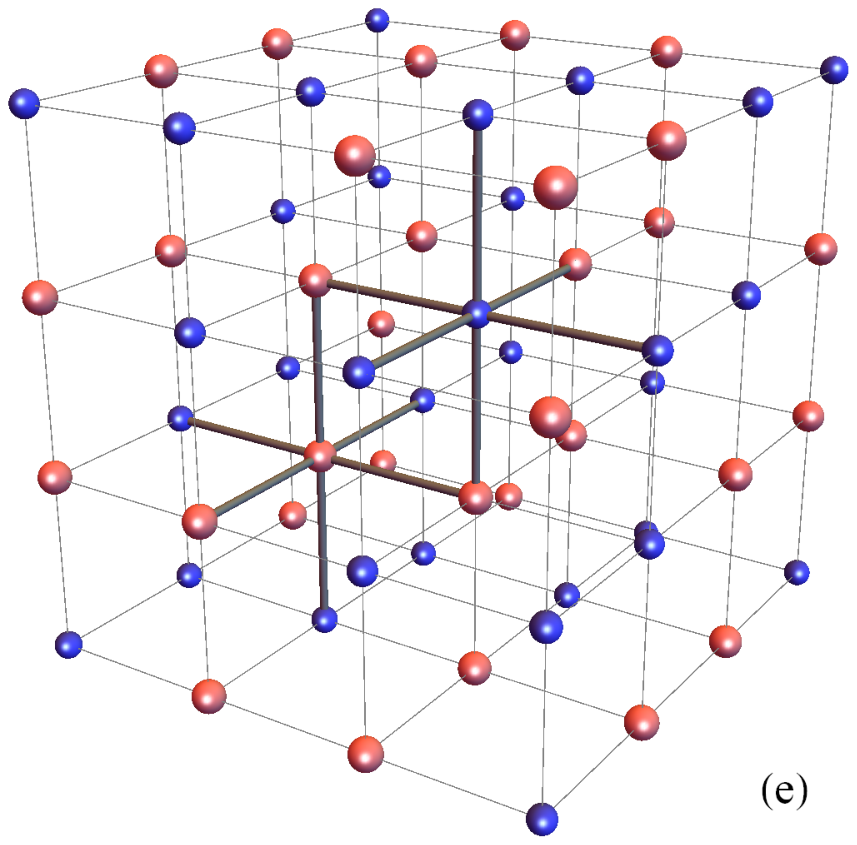}
}
\end{tabular}
\caption{\label{fig:my1} Model of a two-component nuclear system. Blue and red balls are the nuclei of the 1st and the 2nd subsystems, respectively. The lines between the balls indicate the bonds between the nuclei. To make figure (e) more clear some of bonds are not shown. Possible configurations of the position of the nearest neighbourhood nuclei near the "central" nuclear are shown in (a)-(d). The unit cell constructed from the structures (c) and (d) is presented in  (e).}
\end{figure}

Let us consider a two-component nuclear system consisting of inhomogeneously broadened two-level subsystems with central frequencies $\omega_{01}$ and $\omega_{02}$, whose nuclei are coupled by the Ising spin-spin interaction (Fig.~\ref{fig:my0}(b)). We assume the nuclei are uniformly distributed over the sample volume and are located at the cubic lattice knots.
The uniformity of the nuclei distribution can be realized in two ways: 1) nuclei of two types are randomly distributed over the sample volume; 2) nuclei of two types form a periodic structure. In the first case, there are 64 possible configurations for the location of neighboring nuclei near some randomly selected nucleus (“central”). In this case, the most common configuration is in which the central nucleus is surrounded by three nuclei of the first and three nuclei of the second subsystem (Fig.~\ref{fig:my1}(a) - (d)). We consider only these configurations, since in them both the interaction between the subsystems and the interaction within the subsystems are most strongly manifested. In addition, such configurations can, with some assumptions, be considered as part of some larger molecule. From the configurations of the nuclei shown in Fig.~\ref{fig:my1}(c), (d), it is possible to compose a crystal lattice with a unit cell shown in Fig.~\ref{fig:my1}(e). Note that, within the framework of our model, the structure of the Hamiltonian does not depend on the choice of the configuration of the nuclei of the environment. Let's select “central” nucleus. All other nuclei of the lattice affect this “central” nucleus. The spin-spin interaction between nuclei is inversely proportional to the fourth power of the distance between them, therefore the “central” nucleus of the 1st or the 2nd subsystems will be mostly affected only by the nearest nuclei.
Interactions of these six nearest neighbourhood nuclei between each other can be neglected, because the strength of these interactions is ${{\left( \sqrt{2} \right)}^{4}}=4$ times smaller than the strength of interaction with the “central" nucleus. The central nuclei and their neighbours are located in the homogeneous intercrystalline magnetic field.
As it is seen in Fig. \ref{fig:my1} the "central" nuclei of the 1st and 2nd subsystems have the same set of neighbours.

The effective field in which these neighbours are located depends on the behaviour of the "central" nuclei. The different dynamics of the "central" nuclei of the 1st and 2nd subsystems results in the different behaviour of the surrounding nuclei. However, due to the symmetry of the subsystems, to describe the nuclear subsystem dynamics, it is sufficient to investigate the behaviour of one of them. Using \eqref{eq6}, we can write the system of equations for the nuclei magnetization vector components of the 1st subsystem as follows:

\begin{subequations}\allowdisplaybreaks
\label{eq7}
\begin{eqnarray}
\label{eq7a}
{{\dot{u}}_{1}}& = &-{{\upsilon }_{1}}({{\Delta }_{1}}+{{\delta }_{1}}+3{{k}_{1}}{{W}_{1}}+3{{k}_{12}}{{W}_{2}}),\nonumber
\\
{{\dot{\upsilon}}_{1}}&=& -{{\omega }_{R}}{{\eta }_{1}}{{w}_{1}}+{}\nonumber\\
& &{{u}_{1}}({{\Delta }_{1}}+{{\delta }_{1}}+3{{k}_{1}}{{W}_{1}}+3{{k}_{12}}{{W}_{2}}),
\\
{\dot{w}}_{1} &=&{{\omega }_{R}}{{\eta }_{1}}{{\upsilon }_{1}},\nonumber\\
\label{eq7b}
{}\nonumber\\
{{\dot{U}}_{1}} &= &-{{V}_{1}}({{\Delta }_{1}}+{{\delta }_{1}}+{{k}_{1}}{{w}_{1}}),\nonumber
\\
{{\dot{V}}_{1}} &=&{{U}_{1}}({{\Delta }_{1}}+{{\delta }_{1}}+{{k}_{1}}{{w}_{1}})-{{\omega }_{R}}{{\eta }_{1}}{{W}_{1}},
\\
{{\dot{W}}_{1}} &=&{{\omega }_{R}}{{\eta }_{1}}{{V}_{1}},\nonumber\\
{}\nonumber\\
\label{eq7c}
{{\dot{U}}_{2}}&=&-{{V}_{2}}({{\Delta }_{2}}+{{\delta }_{2}}+{{k}_{12}}{{w}_{1}}),\nonumber
\\
{{\dot{V}}_{2}}&=&{{U}_{2}}({{\Delta }_{2}}+{{\delta }_{2}}+{{k}_{12}}{{w}_{1}})-{{\omega }_{R}}{{\eta }_{2}}{{W}_{2}},
\\
{{\dot{W}}_{2}} &=&{{\omega }_{R}}{{\eta }_{2}}{{V}_{2}},\nonumber
\end{eqnarray}
\end{subequations}
where ${{u}_{1}}$, ${{\upsilon }_{1}}$, ${{w}_{1}}$  are the "central" nuclei magnetization vector components of the 1st subsystem, ${{U}_{1,2}}$, ${{V}_{1,2}}$, ${{W}_{1,2}}$ are the corresponding components of surrounding nuclei, ${{\delta }_{1,2}}={{\omega }_{01,02}}-\omega$, ${{\Delta }_{1,2}}={{\omega }_{n1,n2}}-{{\omega }_{01,02}}$, ${{k}_{i}}$ and ${{k}_{12}}$  are the spin-spin coupling constants inside the i-th subsystem (i=1,2) and between the 1st and 2nd subsystems.

We assume that the 1st and the 2nd nuclear subsystems are in thermodynamic equilibrium at the time moment $t=0$. In this case the initial conditions for solving the system of Eqs. \eqref{eq7} can be taken as follows: ${{u}_{1}}(0)={{\upsilon }_{1}}(0)=0$, ${{w}_{1}}(0)=1$, ${{U}_{1,2}}(0)={{V}_{1,2}}(0)=0$, ${{W}_{1,2}}(0)=1$.

The $k_1w_1$ and $k_{12}w_1$ terms in \eqref{eq7b} and \eqref{eq7c} can be neglected because, in frame of our model, the "central" nucleus is affected by six neighbouring nuclei (three nuclei of the 1st subsystem and three nuclei of the 2nd subsystem) and only one "central" nucleus acts on the neighboring nuclei (Fig. \ref{fig:my1}). To confirm the validity of this assumption for analytical calculations, we compared the numerical solutions of \eqref{eq7} with and without the influence of the "central" nucleus on neighbouring nuclei. We found out that for the broad parameter range, the difference between these two solutions does not exceed 1\%. Within this approximation the Eqs. \eqref{eq7b} and \eqref{eq7c} become independent from each other as well as from the Eqs. \eqref{eq7a}. Thus, solving separately equations \eqref{eq7b}, \eqref{eq7c} we obtain
%\left( 1/\beta _{1,2}^{2} \right)
\begin{equation}\label{eq8}
{{W}_{1,2}}=\frac{1}{\beta _{1,2}^{2}} \left( {{({{\Delta }_{1,2}}+{{\delta }_{1,2}})}^{2}}+{{( {{\omega }_{R}}{{\eta }_{1,2}} )}^{2}}\cos ({{\beta }_{1,2}}t) \right),
\end{equation}
where ${{\beta }_{1,2}}=\sqrt{{{\left( {{\omega }_{R}}{{\eta }_{1,2}} \right)}^{2}}+{{({{\Delta }_{1,2}}+{{\delta }_{1,2}})}^{2}}}$ is the generalized Rabi frequency.
Substituting \eqref{eq8} into the system of Eqs. \eqref{eq7a} we obtain a system of linear differential equations with variable coefficients. The following Hamiltonian corresponds to this system:
\begin{equation}\label{eq9}
{{\hat{H}}_{s}}= ({{\Delta }_{1}}+{{\delta }_{1}}+3{{k}_{1}}{{W}_{1}}+3{{k}_{12}}{{W}_{2}})\hat{S}_{1}^{z}+{{\eta }_{1}}{{\omega }_{R}}\hat{S}_{1}^{x}.
\end{equation}

To investigate the behaviour of Hamiltonian \eqref{eq9}, we assume the interaction between the subsystems to be zero. Then we get
\begin{equation}\label{eq10}
\begin{split}
{{\hat{H}}_{s}}=& ({{\Delta }_{1}}+{{\delta }_{1}})\hat{S}_{1}^{z} +{{\eta }_{1}}{{\omega }_{R}}\hat{S}_{1}^{x} +{}\\
&3{{k}_{1}} \frac{({{\Delta }_{1}}+{{\delta }_{1}})^2} {{{\beta}_{1}^{2}}} \hat{S}_{1}^{z} +{}\\
&3{{k}_{1}} \frac{\eta_1^2 \omega_R^2 \cos(\beta_1 t)}{\beta_1^2} \hat{S}_{1}^{z}.
\end{split}
\end{equation}

\begin{figure}[htp]
 \includegraphics[width=8cm]{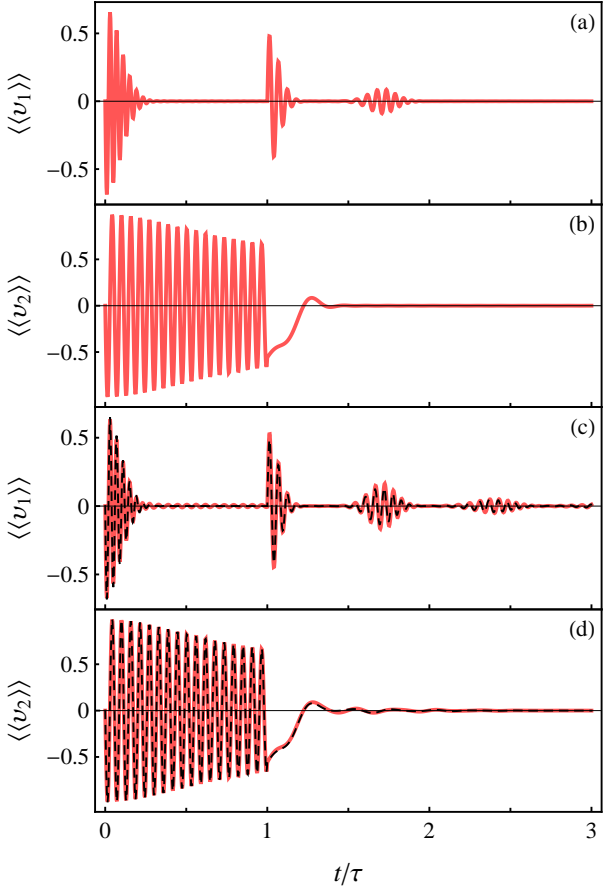}
 \caption{\label{fig:my2} Rabi oscillations ($t< \tau$) and FID ($t>\tau$) in the 1st (a), (c) and 2nd (b), (d) nuclear subsystems for ${{\omega }_{R}}\eta_{1,2} \tau = 34.5\pi$, ${{\delta }_{1}}\tau=113$, ${{\delta }_{2}}\tau =0$, $\sigma\tau =15$, ${{k}_{1}}\tau = {{k}_{2}}\tau={{k}_{12}}\tau=0$ (a), (b); 0.9 (c), (d). The red and dashed black lines are numerical and analytical solutions of equations \eqref{eq7}, respectively. The solutions are obtained with averaging over the form-factor of inhomogeneously broadened lines of each subsystem. The fast oscillations correspond to the generalized Rabi frequencies ${{\beta }_{1,2}}$ for the 1st and 2nd subsystems, respectively.}
\end{figure}

It can be seen from \eqref{eq10} that the third and fourth terms lead to static and time-dependent (dynamic) energy level shifts of the 1st subsystem, respectively \cite{BorovikRomanov.1984,Bunkov.1974,Khasanov.2019,Khasanov.2003}. Note that the Hamiltonian \eqref{eq10} is similar to the one describing a two-level system under parametric excitation \cite{Pokazanev.1977} for which the fourth term is responsible.
Further, we focus on the approximate solution of the system \eqref{eq7a} with variable coefficients ($3{{k}_{1}}{{W}_{1}}$, $3{{k}_{12}}{{W}_{2}}$). These coefficients describe the time-dependent detuning of the "central" nucleus frequency from the carrier frequency of the RF pulse. In our model, we replace this detuning with its average value over the period, expanding the variable coefficients in the cosine Fourier series retaining only the first term of the expansion. A comparison of the analytical solutions obtained below with numerical calculations in (Fig. \ref{fig:my2}) shows that this approach is appropriate and does not lead to any significant deviations from the numerical results. In this approximation the solutions of \eqref{eq7a} are
\begin{eqnarray}
\label{eq11}
{{u}_{1}}(t)&=&\frac{({{\bar{\delta }}_{1}}+{{\Delta }_{1}}){{\eta }_{1}}{{\omega }_{R}}(1-\cos ({{\bar{\beta }}_{1}}t))}{\bar{\beta }_{1}^{2}},\nonumber
\\
{{\upsilon }_{1}}(t)&=&-\frac{{{\omega }_{R}}{{\eta }_{1}}\sin ({{\bar{\beta }}_{1}}t)}{{{\bar{\beta }}_{1}}},
\\
{{w}_{1}}(t)&=&\frac{{{({{\bar{\delta }}_{1}}+{{\Delta }_{1}})}^{2}}+\eta _{1}^{2}\omega _{R}^{2}\cos ({{\bar{\beta }}_{1}}t)}{\bar{\beta }_{1}^{2}},\nonumber
\end{eqnarray}
where ${{\bar{\beta }}_{1}}=\sqrt{{{( {{\omega }_{R}}{{\eta }_{1}} )}^{2}}+{{({{\Delta }_{1}}+{{\bar{\delta }}_{1}})}^{2}}}$ is the effective Rabi frequency, ${{\bar{\delta }}_{1}}={{\delta }_{1}}+3{{k}_{1}}{{\bar{W}}_{1}}+3{{k}_{12}}{{\bar{W}}_{2}}$  is the effective detuning from resonance, ${{\bar{W}}_{1,2}}={{({{\Delta }_{1,2}}+{{\delta }_{1,2}})}^{2}}/\beta _{1,2}^{2}$ are the period-averaged ($2\pi / {{{\beta }}_{1,2}}$) longitudinal magnetization of the 1st and 2nd subsystems of neighboring nuclei. 

Eqs. \eqref{eq11} describe the behaviour of the 1st subsystem nuclei magnetization components during pulse exposure. To find the 1st subsystem component ${{\upsilon }_{1}}$ after the pulse ($t>\tau$) we will solve \eqref{eq7} for ${{\omega }_{R}}=0$. Magnetization components at the end of the RF pulse given by Eqs. \eqref{eq11} can be taken as initial conditions for \eqref{eq7}. In this case \eqref{eq7} has the following analytical solution:
\begin{equation}
\label{eq12}
\begin{split}
{{\upsilon }_{1}}(t)=&{{u}_{1}}(\tau )\sin (({{\Delta }_{1}}+{{\delta }_{1}}+3{{k}_{1}}{{\tilde{W}}_{1}}+
3{{k}_{12}}{{\tilde{W}}_{2}})\tilde t)+{}\\
&{{\upsilon }_{1}}(\tau )\cos (({{\Delta }_{1}}+{{\delta }_{1}}+3{{k}_{1}}{{\tilde{W}}_{1}}+3{{k}_{12}}{{\tilde{W}}_{2}})\tilde{t}),
\end{split}
\end{equation}
where $\tilde{t}=(t-\tau )$, ${{\tilde{W}}_{1,2}}={{W}_{1,2}}(\tau )$.
In our model, the rate of reversible phase relaxation is much higher than the rates of energy and phase irreversible relaxations. Therefore, the inhomogeneous broadening is the only reason for the signal decay. Let's average the component over form-factor of the inhomogeneously broadened lines of the 1st and 2nd subsystems

\begin{equation}
\label{eq13}
\langle \langle {{\upsilon }_{1}} \rangle \rangle ={\iint\limits_{-\infty }^{+\infty }{g({{\Delta }_{1}})g({{\Delta }_{2}})}}{{\upsilon }_{1}}({{\Delta }_{1}},{{\Delta }_{2}})d{{\Delta }_{1}}d{{\Delta }_{2}},
\end{equation}
where $g(x)=1/(\sqrt{2\pi }\sigma)\exp(-{{x}^{2}}/(2{{\sigma }^{2})})$, $\sigma \sqrt{2\ln 2}$ is the half width at half maximum of the inhomogeneously broadened line.

Basing on \cite {BorovikRomanov.1984,Kiliptari.1998,Savosta.2003} we use the following parameters for calculating the Rabi oscillations and the FID in the 1st and 2nd subsystems: the Ising nuclear interaction energy in frequency units is 0.9 MHz, the NMR lines width at half maximum is 30 MHz, the distance between the central frequencies of the nuclear subsystems NMR lines is 113 MHz, the pulse area is $0.345\pi$, the RF pulse duration is 1 $\mu s$, and the RF field gain factor on nuclei is equal to $10^2$. 
We assume that the RF pulse resonantly excites nuclei of the 2nd subsystem. Fig. \ref{fig:my2} shows numerical and analytical solutions of equations \eqref{eq7}. These solutions are averaged over the form-factors of inhomogeneously broadened lines of 1st and 2nd subsystems and are consistent with each other (Figs. \ref{fig:my2}(c), \ref{fig:my2}(d)). It can be seen in Fig. \ref{fig:my2}(a) that in the absence of the Ising interaction in the non-resonantly excited 1st subsystem only one PSPE appears after the initial part of FID (IFID). When the Ising interaction is non-zero, MSPE are generated in the 1st subsystem (Fig. \ref{fig:my2}(c)). No SPE are observed in the 2nd subsystem since under resonant excitation the SPE signals are formed at the time $t= \tau$ simultaneously with the IFID (Fig. \ref{fig:my2}(b) and \ref{fig:my2}(d)). 

Next, we split the contributions of the 1st and 2nd subsystems to the formation of MSPE signals. Since we consider the interaction of nuclear subsystems with a strong RF field (${{k}_{1}}\ll {\eta_1{\omega}_{R}}$, ${{k}_{12}} \ll {\eta_1{\omega}_{R}}$), we can neglect the second and subsequent terms of the Taylor series, while expanding ${{u}_{1}}(\tau)$ and ${{\upsilon}_{1}}(\tau)$ from \eqref{eq12} in a Taylor series by small parameters $k_1$ and $k_{12}$. Substituting \eqref{eq12} into \eqref{eq13}, we get
%\begin{widetext}
\begin{multline}
\label{eq14}
\langle \langle {{\upsilon }_{1}}(t)\rangle \rangle ={{\omega }_{R}}{{\eta }_{1}}\iint_{-\infty }^{+\infty }{\sum\limits_{q=-1}^{1}{}{{F}_{q}}({{\Delta }_{1}})\times{}}
\\
\sin (q{{\beta }_{1}}\tau +({{\Delta }_{1}}+{{\delta}_{1}}+3{{k}_{1}}{{{\tilde{W}}}_{1}}+3{{k}_{12}}{{{\tilde{W}}}_{2}})\tilde{t})\times{}
\\
g({{\Delta }_{1}})g({{\Delta }_{2}})d{{\Delta }_{1}}d{{\Delta }_{2}},
\end{multline}
%\end{widetext}
where 
\begin{equation*}
F_q(\Delta _1) = \frac{(-1)^q}{1+q^2}\frac{q \beta _1+\delta _1+\Delta _1}{\beta _1^2}.
\end{equation*}

Let us rewrite \eqref{eq14} in a more convenient form:
\begin{multline}
\label{eq15}
\langle \langle \upsilon _1(t)\rangle \rangle = \omega _R\eta_1 \sum _{q=-1}^1\\ 
%(
\iint_{-\infty }^{+\infty }{}({{F}_{q}}({{\Delta }_{1}})\sin (q{{\beta }_{1}}\tau +({{\Delta }_{1}}+{{\delta }_{1}}+3{{k}_{1}}{{\tilde{W}}_{1}})\tilde{t})\times{}\\\cos (3{{k}_{12}}{{\tilde{W}}_{2}}\tilde{t}))g({{\Delta }_{1}})g({{\Delta }_{2}})d{{\Delta }_{1}}d{{\Delta }_{2}}+{}\\ 
\iint_{-\infty }^{+\infty }{}({{F}_{q}}({{\Delta }_{1}})\cos (q{{\beta }_{1}}\tau +({{\Delta }_{1}}+{{\delta }_{1}}+3{{k}_{1}}{{\tilde{W}}_{1}})\tilde{t})\times{}\\\sin (3{{k}_{12}}{{\tilde{W}}_{2}}\tilde{t}))g({{\Delta }_{1}})g({{\Delta }_{2}})d{{\Delta }_{1}}d{{\Delta }_{2}}
%)
\end{multline}

Since each of the double integral factors \eqref{eq15} depends only on one integration variable, these integrals can be rewritten as the product of ordinary integrals
\begin{multline}     
\label{eq16}
\langle \langle {{\upsilon }_{1}}(t)\rangle \rangle ={{\omega }_{R}}{{\eta }_{1}}\times{}
\\
\sum\limits_{q=-1}^{1}{}(\int_{-\infty }^{+\infty }{}\cos(3{{k}_{12}}{{{\tilde{W}}}_{2}}\tilde{t})g({{\Delta }_{2}})d{{\Delta }_{2}}\times{}\\
\int_{-\infty }^{+\infty }{}({{F}_{q}}({{\Delta }_{1}})\sin (q{{\beta }_{1}}\tau +({{\Delta }_{1}}+{{\delta }_{1}} + 3{{k}_{1}}{{\tilde{W}}_{1}})\tilde{t}))g({{\Delta }_{1}})d{{\Delta }_{1}}+{}
\\ 
+\int_{-\infty }^{+\infty }{}\sin(3{{k}_{12}}{{\tilde{W}}_{2}}\tilde{t})g({{\Delta }_{2}})d{{\Delta }_{2}}\times{}
\\
\int_{-\infty }^{+\infty }{({{F}_{q}}(}{{\Delta }_{1}})\cos (q{{\beta }_{1}}\tau +({{\Delta }_{1}}+{{\delta }_{1}} + 3{{k}_{1}}{{\tilde{W}}_{1}})\tilde{t}))g({{\Delta }_{1}})d{{\Delta }_{1}}).
\end{multline}

It is expedient now to represent Eq. \eqref{eq16} as a product of two functions
\begin{equation}
\label{eq17}
\langle \langle {{\upsilon }_{1}}(t)\rangle \rangle = {\bar A}_2 {\bar A}_1,
\end{equation}where
\begin{eqnarray}
{{\bar{A}}_{1}}&=&{{\eta }_{1}}{{\omega }_{R}}\sum\limits_{q=-1}^{1}{}\int_{-\infty }^{+\infty }{}{{F}_{q}}({{\Delta }_{1}})g({{\Delta }_{1}})\times{}\nonumber\\
\label{eq18}
& &\sin (q{{\beta }_{1}}\tau +({{\Delta }_{1}}+{{\delta }_{1}}+3{{k}_{1}}{{\tilde{W}}_{1}})\tilde{t}+{{\phi }_{2}})d{{\Delta }_{1}},
\\
\label{eq19}
{{\bar{A}}_{2}}&=&\left| \int_{-\infty }^{+\infty }{\exp }(i3{{k}_{12}}{{{\tilde{W}}}_{2}}\tilde{t})g({{\Delta }_{2}})d{{\Delta }_{2}} \right|,
\\
\label{eq20}
{{\phi }_{2}}&=&\arg (\int_{-\infty }^{+\infty }{\exp }(i3{{k}_{12}}{{\tilde{W}}_{2}}\tilde{t})g({{\Delta }_{2}})d{{\Delta }_{2}}).
\end{eqnarray}
\begin{figure}[htp]
 \includegraphics[width=8cm]{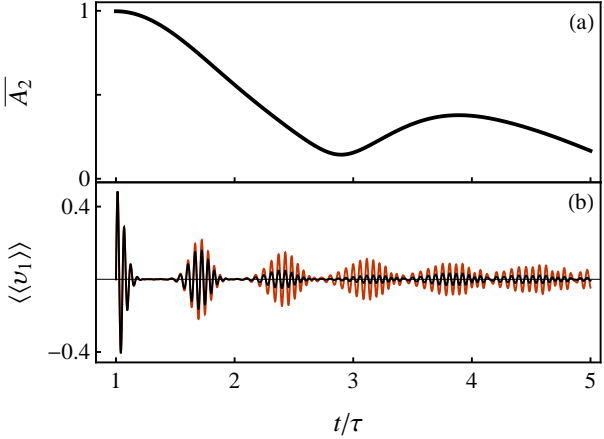}

 \caption{\label{fig:my3} Influence of the Ising interaction on FID: (a) function modulating the FID in the 1st non-resonantly excited subsystem, (b) FID with (black line) and without (red line) taking into account the influence of amplitude modulation; ${{\omega }_{R}}\eta_1 \tau =34.5\pi$, ${{k}_{1}} \tau ={{k}_{12}} \tau =0.9$, ${{\delta }_{1}}\tau =113$, ${{\delta }_{2}} \tau =0$, $\sigma \tau =15$.}

\end{figure}

Note, that in Eq. \eqref{eq17} the influence of the resonantly excited subsystem on the FID of the non-resonantly excited one is described by functions \eqref{eq19} and \eqref{eq20}. These functions oscillate with a frequency (depending on $3k_{12}$ value) which is much smaller than the FID oscillation frequency in the 1st subsystem. This leads to the amplitude and phase modulation of this signal (Fig. \ref{fig:my3}).
It can be seen from Fig. \ref{fig:my3} that the considered modulations cause an non-monotonic attenuation of the MSPE amplitude. When the Ising interaction between resonantly and non-resonantly excited subsystems is zero, functions \eqref{eq19} and \eqref{eq20} are equal to unity and zero, respectively. Therefore, there are no amplitude and phase modulations and the SPE amplitudes decay monotonically.
Let us consider in more detail the behavior of the factor $A_1$ in \eqref{eq17}. It is seen from \eqref{eq18} that this factor does not depend on ${{k}_{12}}$ and describes the first spin subsystem without the influence of the second one. But the behavior of spins inside the first subsystem is described by the $\left\langle\left\langle{{\upsilon}_{1}}\right\rangle\right\rangle$, which depends on $A_2$ and, therefore, on the magnitude of the spin-spin interaction between subsystems. Factor $A_1$ in \eqref{eq17} can be expanded in a series as follows \cite{Khasanov.2003, Kuzmin.2001b}:
\begin{multline}
 \label{eq21}
 \langle\langle \upsilon _1(t)\rangle\rangle = (\bar{A}_2 \omega _R \eta_1 )\times{}
 \\
 \int_{-\infty }^{+\infty } (\sum _{q=-1}^1 \frac{(-1)^q}{1+q^2}\frac{q \beta _1+\delta _1+\Delta _1}{\beta _1^2}\xi _q) g(\Delta _1) \, d\Delta _1,
\end{multline}
where
\begin{eqnarray*}
{{\xi }_{q}}&=&{{J}_{0}}({{\theta }_{1}}\tilde{t})\sin (q{{\beta }_{1}}\tau +{{\alpha }_{1}}\tilde{t}+{{\phi }_{2}})+{}
\\
2&\sum\limits_{n=1}^{\infty }&{{{J}_{n}}}({{\theta }_{1}}\tilde{t})\cos (n{{\beta }_{1}}\tau )\sin (q{{\beta }_{1}}\tau +{{\alpha }_{1}}\tilde{t}+{{\phi }_{2}}+\frac{\pi n}{2}),
\\
{{\alpha }_{1}}&=&({{\delta }_{1}}+{{\Delta }_{1}}+3{{k}_{1}}{{({{\Delta }_{1}}+{{\delta }_{1}})}^{2}}/\beta _{1}^{2}),
\\
{{\theta }_{1}}&=&\frac{3{{k}_{1}}\omega _{R}^{2}}{\beta _{1}^{2}},
\end{eqnarray*}

Let us rewrite Eq. \eqref{eq21} as follows:
\begin{equation}
\label{eq22}
\langle \langle \upsilon _1(t)\rangle\rangle = \bar{A}_2 \eta_1 \omega _R \sum _{n=0}^{\infty } \langle \epsilon _n \rangle,
\end{equation}
where
\begin{multline}
\label{eq23}
\langle {{\epsilon }_{0}}\rangle =\int_{-\infty }^{+\infty }{\frac{({{\delta }_{1}}+{{\Delta }_{1}})}{\beta _{1}^{2}}}g({{\Delta }_{1}})\times{}
\\
({{J}_{0}}({{\theta }_{1}}\tilde{t})\sin ({{\alpha }_{1}}\tilde{t}+{{\phi }_{2}})-
{{J}_{1}}({{\theta }_{1}}\tilde{t})\cos ({{\alpha }_{1}}\tilde{t}+{{\phi }_{2}}))d{{\Delta }_{1}},
\end{multline}
\begin{equation}
\label{eq24}
 \langle \epsilon _n\rangle = \langle \epsilon _n^{+ }\rangle + \langle \epsilon _n^{- }\rangle, (n>0),
\end{equation}
\begin{equation}
\label{eq25}
 \langle \epsilon _{n}^{\pm }\rangle =\int_{-\infty }^{+\infty }{|\zeta _{n}^{\pm }({{\Delta }_{1}})|\sin (\psi _{n}^{\pm }({{\Delta }_{1}}))g({{\Delta }_{1}})d{{\Delta }_{1}}},
\end{equation}
\begin{equation}
\label{eq26}
 \psi _{n}^{\pm }({{\Delta }_{1}})={{\alpha }_{1}}\tilde{t}+{{\phi }_{2}}\pm {{\beta }_{1}}n\tau +\frac{\pi n}{2}+\arg (\zeta _{n}^{\pm }({{\Delta }_{1}})),
\end{equation}
\begin{multline}
\label{eq27}
 \zeta _{n}^{\pm }(\Delta_1) = \beta _{1}^{-2}(({{\Delta }_{1}}+{{\delta }_{1}}){{J}_{n}}({{\theta }_{1}}\tilde{t})+{}\\
 \frac{i}{2}(({{\Delta }_{1}}+{{\delta }_{1}}\pm {{\beta }_{1}}){{J}_{n-1}}({{\theta }_{1}}\tilde{t})-{}
 \\({{\Delta }_{1}}+{{\delta }_{1}}\mp {{\beta }_{1}}){{J}_{1+n}}({{\theta }_{1}}\tilde{t}))).
\end{multline}

Let us estimate the integrals \eqref{eq26} approximately using the stationary phase method \cite{Nayfeh.2011,Kuzmin.1990}
\begin{multline}
\label{eq28}
 \langle \epsilon _{n}^{\pm }\rangle =\frac{\sqrt{2\pi }|\zeta _{n}^{\pm }(\Delta _{1}^{\pm })|}{\sqrt{|(\psi _{n}^{\pm }(\Delta _{1}^{\pm }))^{\prime\prime}|}}g(\Delta _{1}^{\pm })\times{}
 \\
 \sin (\psi _{n}^{\pm }(\Delta _{1}^{\pm })+\mathrm{sgn} (\psi _{n}^{\pm }(\Delta _{1}^{\pm }))\frac{\pi }{4}),
\end{multline}
where ${{\Delta}_{1}^{\pm}} = \pm \omega_R \eta_1 \tilde t/ \sqrt{ n^2 \tau^2- { \tilde t}^2} - \delta_1$ are the stationary phase points. The formulas for these stationary points are obtained in the approximation when $k_1 \ll \eta_1 \omega_R$. We estimate the integral \eqref{eq23} by expanding the slowly varying part of the integrand into a Taylor series up to order zero at the point where the function $g(\Delta_1)$ has its maximum value
\begin{multline}
\langle {{\epsilon }_{0}}\rangle = \frac{1}{\beta _1^2}((-\delta _1J_1(\theta _1 \tilde{t})+\sigma ^2J_0(\theta _1 \tilde{t})\tilde {t})\cos ({{\alpha }_{1}}\tilde{t}+{{\phi }_{2}})+{}
\\
(\delta _1J_0(\theta _1 \tilde{t})+\sigma ^2J_1(\theta _1 \tilde{t})\tilde {t})\sin ({{\alpha }_{1}}\tilde{t}+{{\phi }_{2}}))e^{-\frac{1}{2} \sigma ^2{\tilde {t}}^2}, 
\end{multline}
where $\Delta_1 = 0$.
\begin{figure}[htp]
 \includegraphics[width=8cm]{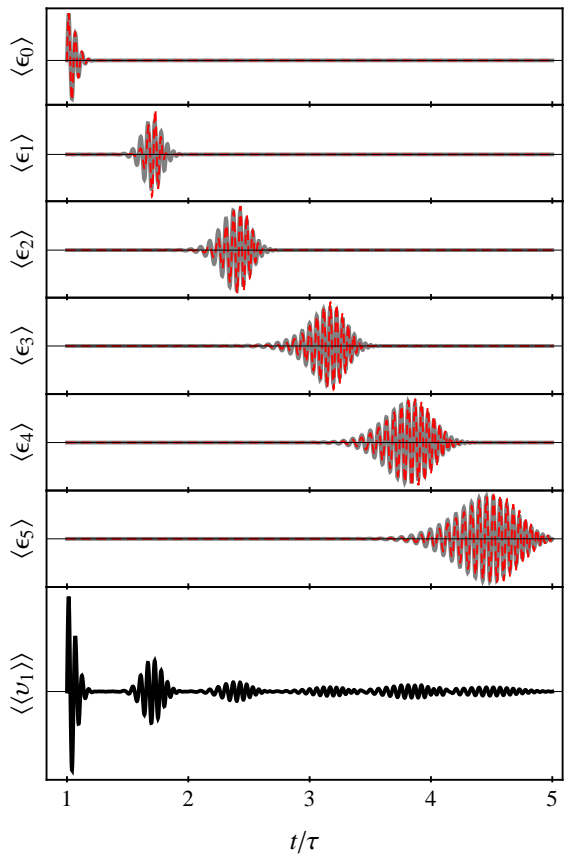}
 \caption{\label{fig:my4} FID in a nuclear subsystem non-resonantly excited by an RF pulse (bottom graph) and its components (six top graphs, normalized to unity). The gray and red dashed lines correspond to numerical and analytical calculations according to formulas \eqref{eq27} and \eqref{eq28}, respectively. The upper graph is the component representing the IFID. The rest of the components are the n-th signals of the SPE in the FID signal $\langle{ \epsilon_n } \rangle$ (${n=\overline{1,5}}$): ${{\omega }_{R}}\eta_1 \tau =34.5\pi$, ${{k}_{1}} \tau ={{k}_{12}} \tau =0.9$, ${{\delta }_{1}}\tau =0$, ${{\delta }_{2}} \tau =113$, $\sigma \tau =15$.}
\end{figure}

 The first term of Eq. \eqref{eq22} describes IFID (Eq. \eqref{eq23}). The subsequent terms \eqref{eq24} are responsible for MSPE with a multiplicity that is equal to the index of this term (Fig. \ref{fig:my4}).
One can see from \eqref{eq24} that the term describing the n-th echo signal is a sum of two quantities, $\langle\epsilon^{+}\rangle$ and $\langle {{\epsilon }^{-}} \rangle $, which have a similar structure \eqref{eq25}. However, estimations by the stationary phase method \eqref{eq28} show that only the second term contributes to the observed echo signal. From the formula for this term we find the moment of n-th SPE signal formation in the approximation when $k_1 \ll \eta_1 \omega_R$ \cite{Kuzmin.1990,Kuzmin.2001} 
\begin{equation}
t_n= n\tau\delta_1\frac{\beta_1^3}{\eta_1^2\omega_R^2\delta_1 k_1 +\beta_1^4}.
 %t_n = n\tau\delta_1\frac{\beta_1}^3{\omega_R^2\delta_1 k_1 +\beta_1^4}
\end{equation}

Thus, we have obtained an approximate analytical description of the Rabi oscillations and the FID generated in the two-component nuclear system with spin-spin interaction. We have derived formulas for calculating the shape, amplitude and moments of formation of the PSPE and MSPE.
\section{Discussion}

Let us consider in detail the effects to which the Ising interaction leads in a two-component nuclear system exposed to a single pulse. We analyze first the Rabi oscillations. It's known that the decay rate of the Rabi oscillations of an inhomogeneously broadened two-level system grows with the increase in the detuning of the RF field frequency from resonance \cite{Kuzmin.2001b}. In our case, the 1st subsystem is excited non-resonantly. Therefore, the Rabi oscillations decay is faster in the 1st subsystem than in the 2nd resonantly excited subsystem (Fig. \ref{fig:my2}). 

\begin{figure}[htp]
 \includegraphics[width=8cm]{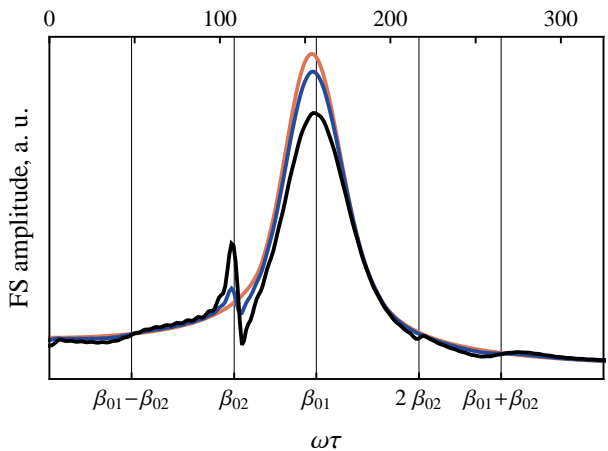}
 \caption{\label{fig:my5} Fourier spectrum of Rabi oscillations of the 1st nuclear subsystem for ${{k}_{1}} \tau ={{k}_{2}} \tau = {{k}_{12}} \tau = 0$ (red line), 0.9 (blue line), 3 (black line); ${{\omega }_{R}}{{\eta }_{1,2}} \tau =34.5\pi$, ${{\delta }_{1}} \tau =113$, ${{\delta }_{2}} \tau = 0$, $\sigma \tau =15$, ${{\beta }_{01,02}}=\sqrt{{{\left( {{\omega }_{R}}{{\eta }_{1,2}} \right)}^{2}}+{{\delta }_{1,2}}^{2}}$. }
\end{figure}

The spectrum of Rabi oscillations of the two-component system in the absence of the Ising interaction consists of narrow and wide lines with frequencies ${{\beta }_{01}}$ and ${{\beta }_{02}}$. The narrow line corresponds to the resonantly excited subsystem. The interaction between subsystems and nuclei inside each subsystem affects the response of the two-component system. Since the nuclei of the 2nd subsystem are stronger excited (resonant excitation), their impact on the 1st subsystem is more significant than the reverse effect. Therefore, we will further consider the influence of the spin-spin interaction on the 1st subsystem. With an increase of the interaction strength weak and slowly damped oscillations appear in the Rabi oscillations of the 1st subsystem (Fig. \ref{fig:my2}(a)). These oscillations are due to the influence of the resonantly excited subsystem. They appear in the spectrum of the non-resonant subsystem as a weak and narrow line at the frequency ${{\beta }_{02}}$ (Fig. \ref{fig:my5}). The intensity of this spectral line increases with the growth of the interaction coefficient (Fig. \ref{fig:my5}).
For the case of strong nonlinear interaction (value of ${{k}_{12}\tau}\gg 1$) the Rabi oscillations begin to include a harmonic at the doubled Rabi frequency $2{{\beta }_{02}}$ and lines with frequencies ${{\beta }_{01}} \pm {{\beta }_{02}}$. The harmonic at the frequency $2{{\beta }_{02}}$ indicates the anharmonicity of the Rabi oscillations in the second nuclear subsystem. This anharmonicity is caused by the interaction of nuclei inside the second subsystem. The lines with the frequencies ${{\beta }_{01}} \pm {{\beta }_{02}}$ are due to the nonlinear interaction of the nuclei of the first and second subsystems. Such features of the spectrum make it possible to estimate the strength of the spin-spin interaction between the nuclear subsystems of two-component nuclear spin system. The increase in the interaction strength, causing quite small effects in the Rabi oscillation, leads to the generation of MSPE in the FID (Fig. \ref{fig:my2}(c)). Spin-spin interaction is a necessary condition for the MSPE appearance. As noted above, this interaction leads to static and dynamic shifts of the nuclear subsystems transition frequency.

With a growth of the Ising interaction the MSPE amplitude first increases, reaches its maximum and then decreases. Such behaviour of the MSPE amplitude is due to the growing influence of the 2nd subsystem on the 1st one.

The spins of nuclei inside the i-th subsystem oscillate with frequencies close to $ {{\beta} _ {0i}} $.  The oscillations of a selected nucleus of the 1st subsystem (due to the spin-spin interaction) strongly depend on the influence of the nearest neighbours. Nuclei with close oscillation frequencies interact more efficiently. Thus, the influence on the selected nucleus of the 1st subsystem neighbours nuclei is stronger than the impact of the 2nd one.

Therefore, if a spin-spin coupling constant is small, the observed signal from the first subsystem is mainly caused by this subsystem nuclei response. The interaction between the nuclei of the 1st subsystem leads to the formation of multiple echoes, whose amplitudes increase with the Ising interaction growth (Fig. \ref{fig:my2} (a), (c)). Further increase of the spin-spin interaction constant leads to the higher impact of the 2nd subsystem on the 1st one. This impact leads to the amplitude modulation of the multiple echo responses (Fig. \ref{fig:my3}).

At the used approximation  (${{k}_{1}}\ll{\eta_1 {\omega}_{R}}$, ${{k}_{12}}\ll{\eta_1 {\omega}_{R}}$) the frequency distributions of the nuclei magnetization components are practically identical in cases of absence and presence of spin-spin interactions. Thus, after the averaging \eqref{eq13}, the behaviour of the Rabi oscillations (Fig. 2(a) and 2(c)) is similar. The spin-spin interaction during the pulse exposure causes a weak nonlinearity of magnetization components but does not lead to the formation of MSPE. However, due to the spin-spin interaction, the FID starts to contain information on the longitudinal magnetization components of the neighbouring nuclei $W_1(\tau)$ at the end of the pulse. The presence of this information is the reason for the multiple echoes generation.

One can see from \eqref{eq22} that PSPE and MSPE signals in the 1st subsystem are caused by the nonlinear dependence of the phases \eqref{eq26} of the induction signal terms with $n>0$ on the detuning $\Delta_1$. This non-linearity is described by the term $n \beta_1 \tau$. This term depends on the nucleus and its neighbouring nuclei phases of magnetization at the end of the pulse exposure.
In the absence of the Ising interaction the FID has only one term ($\epsilon_1$) with a nonlinear phase. This term describes PSPE (Fig. \ref{fig:my2}(a)). Other terms with $ n> 1$ are equal to zero since $ \theta=0 $ and $ J_ {m} (0) = 0$ if $m\geq1$.

When the spin-spin interaction is not equal to zero one also gets terms with $n>1$ describing MSPE in the FID (Fig. \ref{fig:my2}(c)). These echoes arise from the influence of the environment on the "central" nucleus. The PSPE generation is due to the "central" and neighbouring nuclei.
Thus, it is possible to separately estimate the contributions of a nucleus and its environment to the FID by analyzing the PSPE and MSPE. Moreover, the results on MSPE allow us to predict the behaviour of the widely used two-pulse echo \cite{Berzhanskii.2020}. This is possible because the two-pulse echo contains a time-inverted FID after the first pulse \cite{Kuzmin.2002, Saiko.1989}. However, Hahn's echo has a quite complicated multi-peak structure even in one-component systems \cite{Saiko.1990, Berzhanskii.2020}. If we consider a two-component system with a spin-spin interaction our results predict that this structure becomes more complicated and it can provide the same information on the spin-spin interaction as the one from the MSPE. Since the amount of peaks in Hahn's echo would be significantly larger than that in the FID the interpretation of the former would be much more difficult. Therefore, to our sight, the use of a SPE to obtain information on strength of spin-spin interaction from multi-component systems can be more preferable in a lot of cases.
\section{Conclusions}
We have obtained an analytical description of Rabi oscillations and free induction decay, which are formed in two-component nuclear systems in manganites, taking into account the Ising interaction of nuclei. We investigated these signals when one subsystem (the first one) was excited non-resonantly, and the second one was excited resonantly by a radio-frequency pulse. We have shown the Rabi oscillations of the first subsystem have additional harmonics caused by the second subsystem. The frequencies of these harmonics are multiples of the oscillation frequency of the second subsystem. The amplitude of harmonics depends on the nuclei interaction strength between the subsystems and allows one to estimate the spin-spin interaction constant. We have found the FID magnetization components of the first subsystem contain information on the longitudinal magnetization of the nearest neighbours at the end of the RF pulse. This peculiarity leads to nonlinearity in a phase of the FID term, and it is a reason for the MSPE generation. We also have found that the interaction of nuclei between subsystems causes additional decay of MSPE. We have demonstrated that the primary and multiple signals of single-pulse echoes can allow one to separately estimate the nucleus and its surroundings contributions to the FID. This knowledge provides additional spectroscopic information on two-component nuclear systems with spin-spin interactions.
%\section{Acknowledgments}
%The authors are grateful for the valuable %advice while preparing the paper to R. %Fedaruk, A.P. Saiko and O.M. Fedotova.

%large figure 2
%\begin{figure*}
%\includegraphics{fig_2}% Here is how to import EPS art
%\caption{\label{fig:wide}Use the figure* environment to get a wide
%figure that spans the page in \texttt{twocolumn} formatting.}
%\end{figure*}

\bibliographystyle{apsrev4-2}
\bibliography{1PhysRevB}

\begin{thebibliography}{49}%
\makeatletter
\providecommand \@ifxundefined [1]{%
 \@ifx{#1\undefined}
}%
\providecommand \@ifnum [1]{%
 \ifnum #1\expandafter \@firstoftwo
 \else \expandafter \@secondoftwo
 \fi
}%
\providecommand \@ifx [1]{%
 \ifx #1\expandafter \@firstoftwo
 \else \expandafter \@secondoftwo
 \fi
}%
\providecommand \natexlab [1]{#1}%
\providecommand \enquote  [1]{``#1''}%
\providecommand \bibnamefont  [1]{#1}%
\providecommand \bibfnamefont [1]{#1}%
\providecommand \citenamefont [1]{#1}%
\providecommand \href@noop [0]{\@secondoftwo}%
\providecommand \href [0]{\begingroup \@sanitize@url \@href}%
\providecommand \@href[1]{\@@startlink{#1}\@@href}%
\providecommand \@@href[1]{\endgroup#1\@@endlink}%
\providecommand \@sanitize@url [0]{\catcode `\\12\catcode `\$12\catcode
  `\&12\catcode `\#12\catcode `\^12\catcode `\_12\catcode `\%12\relax}%
\providecommand \@@startlink[1]{}%
\providecommand \@@endlink[0]{}%
\providecommand \url  [0]{\begingroup\@sanitize@url \@url }%
\providecommand \@url [1]{\endgroup\@href {#1}{\urlprefix }}%
\providecommand \urlprefix  [0]{URL }%
\providecommand \Eprint [0]{\href }%
\providecommand \doibase [0]{https://doi.org/}%
\providecommand \selectlanguage [0]{\@gobble}%
\providecommand \bibinfo  [0]{\@secondoftwo}%
\providecommand \bibfield  [0]{\@secondoftwo}%
\providecommand \translation [1]{[#1]}%
\providecommand \BibitemOpen [0]{}%
\providecommand \bibitemStop [0]{}%
\providecommand \bibitemNoStop [0]{.\EOS\space}%
\providecommand \EOS [0]{\spacefactor3000\relax}%
\providecommand \BibitemShut  [1]{\csname bibitem#1\endcsname}%
\let\auto@bib@innerbib\@empty
%</preamble>
\bibitem [{\citenamefont {Zangara}\ \emph {et~al.}(2015)\citenamefont
  {Zangara}, \citenamefont {Bendersky},\ and\ \citenamefont
  {Pastawski}}]{Zangara.2015}%
  \BibitemOpen
  \bibfield  {author} {\bibinfo {author} {\bibfnamefont {P.~R.}\ \bibnamefont
  {Zangara}}, \bibinfo {author} {\bibfnamefont {D.}~\bibnamefont {Bendersky}},\
  and\ \bibinfo {author} {\bibfnamefont {H.~M.}\ \bibnamefont {Pastawski}},\
  }\href {https://doi.org/10.1103/PhysRevA.91.042112} {\bibfield  {journal}
  {\bibinfo  {journal} {Phys. Rev. A}\ }\textbf {\bibinfo {volume} {91}},\
  \bibinfo {pages} {042112} (\bibinfo {year} {2015})}\BibitemShut {NoStop}%
\bibitem [{\citenamefont {Zangara}\ and\ \citenamefont
  {Pastawski}(2017)}]{Zangara.2017}%
  \BibitemOpen
  \bibfield  {author} {\bibinfo {author} {\bibfnamefont {P.~R.}\ \bibnamefont
  {Zangara}}\ and\ \bibinfo {author} {\bibfnamefont {H.~M.}\ \bibnamefont
  {Pastawski}},\ }\href {https://doi.org/10.1088/1402-4896/aa5bee} {\bibfield
  {journal} {\bibinfo  {journal} {Physica Scripta}\ }\textbf {\bibinfo {volume}
  {92}},\ \bibinfo {pages} {033001} (\bibinfo {year} {2017})}\BibitemShut
  {NoStop}%
\bibitem [{\citenamefont {Kaur}\ \emph {et~al.}(2013)\citenamefont {Kaur},
  \citenamefont {Ajoy},\ and\ \citenamefont {Cappellaro}}]{Kaur.2013}%
  \BibitemOpen
  \bibfield  {author} {\bibinfo {author} {\bibfnamefont {G.}~\bibnamefont
  {Kaur}}, \bibinfo {author} {\bibfnamefont {A.}~\bibnamefont {Ajoy}},\ and\
  \bibinfo {author} {\bibfnamefont {P.}~\bibnamefont {Cappellaro}},\ }\href
  {https://doi.org/10.1088/1367-2630/15/9/093035} {\bibfield  {journal}
  {\bibinfo  {journal} {New Journal of Physics}\ }\textbf {\bibinfo {volume}
  {15}},\ \bibinfo {pages} {093035} (\bibinfo {year} {2013})}\BibitemShut
  {NoStop}%
\bibitem [{\citenamefont {S{\'a}nchez}\ \emph {et~al.}(2014)\citenamefont
  {S{\'a}nchez}, \citenamefont {Acosta}, \citenamefont {Levstein},
  \citenamefont {Pastawski},\ and\ \citenamefont {Chattah}}]{Sanchez.2014}%
  \BibitemOpen
  \bibfield  {author} {\bibinfo {author} {\bibfnamefont {C.~M.}\ \bibnamefont
  {S{\'a}nchez}}, \bibinfo {author} {\bibfnamefont {R.~H.}\ \bibnamefont
  {Acosta}}, \bibinfo {author} {\bibfnamefont {P.~R.}\ \bibnamefont
  {Levstein}}, \bibinfo {author} {\bibfnamefont {H.~M.}\ \bibnamefont
  {Pastawski}},\ and\ \bibinfo {author} {\bibfnamefont {A.~K.}\ \bibnamefont
  {Chattah}},\ }\href {https://doi.org/10.1103/PhysRevA.90.042122} {\bibfield
  {journal} {\bibinfo  {journal} {Physical Review A}\ }\textbf {\bibinfo
  {volume} {90}},\ \bibinfo {pages} {042122} (\bibinfo {year}
  {2014})}\BibitemShut {NoStop}%
\bibitem [{\citenamefont {Guerry}\ \emph {et~al.}(2017)\citenamefont {Guerry},
  \citenamefont {Brown},\ and\ \citenamefont {Smith}}]{Guerry.2017}%
  \BibitemOpen
  \bibfield  {author} {\bibinfo {author} {\bibfnamefont {P.}~\bibnamefont
  {Guerry}}, \bibinfo {author} {\bibfnamefont {S.~P.}\ \bibnamefont {Brown}},\
  and\ \bibinfo {author} {\bibfnamefont {M.~E.}\ \bibnamefont {Smith}},\ }\href
  {https://doi.org/10.1016/j.jmr.2017.08.006} {\bibfield  {journal} {\bibinfo
  {journal} {Journal of magnetic resonance}\ }\textbf {\bibinfo {volume}
  {283}},\ \bibinfo {pages} {22} (\bibinfo {year} {2017})}\BibitemShut
  {NoStop}%
\bibitem [{\citenamefont {Morgan}\ \emph {et~al.}(2012)\citenamefont {Morgan},
  \citenamefont {Oganesyan},\ and\ \citenamefont {Boutis}}]{Morgan.2012}%
  \BibitemOpen
  \bibfield  {author} {\bibinfo {author} {\bibfnamefont {S.~W.}\ \bibnamefont
  {Morgan}}, \bibinfo {author} {\bibfnamefont {V.}~\bibnamefont {Oganesyan}},\
  and\ \bibinfo {author} {\bibfnamefont {G.~S.}\ \bibnamefont {Boutis}},\
  }\href {https://doi.org/10.1103/PhysRevB.86.214410} {\bibfield  {journal}
  {\bibinfo  {journal} {Physical review. B}\ }\textbf {\bibinfo {volume} {86}},\ \bibinfo {pages} {214410}
  (\bibinfo {year} {2012})}\BibitemShut {NoStop}%
\bibitem [{\citenamefont {{\'A}lvarez}\ \emph {et~al.}(2015)\citenamefont
  {{\'A}lvarez}, \citenamefont {Suter},\ and\ \citenamefont
  {Kaiser}}]{Alvarez.2015}%
  \BibitemOpen
  \bibfield  {author} {\bibinfo {author} {\bibfnamefont {G.~A.}\ \bibnamefont
  {{\'A}lvarez}}, \bibinfo {author} {\bibfnamefont {D.}~\bibnamefont {Suter}},\
  and\ \bibinfo {author} {\bibfnamefont {R.}~\bibnamefont {Kaiser}},\ }\href
  {https://doi.org/10.1126/science.1261160} {\bibfield  {journal} {\bibinfo
  {journal} {Science (New York, N.Y.)}\ }\textbf {\bibinfo {volume} {349}},\
  \bibinfo {pages} {846} (\bibinfo {year} {2015})}\BibitemShut {NoStop}%
\bibitem [{\citenamefont {Bernien}\ \emph {et~al.}(2017)\citenamefont
  {Bernien}, \citenamefont {Schwartz}, \citenamefont {Keesling}, \citenamefont
  {Levine}, \citenamefont {Omran}, \citenamefont {Pichler}, \citenamefont
  {Choi}, \citenamefont {Zibrov}, \citenamefont {Endres}, \citenamefont
  {Greiner}, \citenamefont {Vuleti{\'c}},\ and\ \citenamefont
  {Lukin}}]{Bernien.2017}%
  \BibitemOpen
  \bibfield  {author} {\bibinfo {author} {\bibfnamefont {H.}~\bibnamefont
  {Bernien}}, \bibinfo {author} {\bibfnamefont {S.}~\bibnamefont {Schwartz}},
  \bibinfo {author} {\bibfnamefont {A.}~\bibnamefont {Keesling}}, \bibinfo
  {author} {\bibfnamefont {H.}~\bibnamefont {Levine}}, \bibinfo {author}
  {\bibfnamefont {A.}~\bibnamefont {Omran}}, \bibinfo {author} {\bibfnamefont
  {H.}~\bibnamefont {Pichler}}, \bibinfo {author} {\bibfnamefont
  {S.}~\bibnamefont {Choi}}, \bibinfo {author} {\bibfnamefont {A.~S.}\
  \bibnamefont {Zibrov}}, \bibinfo {author} {\bibfnamefont {M.}~\bibnamefont
  {Endres}}, \bibinfo {author} {\bibfnamefont {M.}~\bibnamefont {Greiner}},
  \bibinfo {author} {\bibfnamefont {V.}~\bibnamefont {Vuleti{\'c}}},\ and\
  \bibinfo {author} {\bibfnamefont {M.~D.}\ \bibnamefont {Lukin}},\ }\href
  {https://doi.org/10.1038/nature24622} {\bibfield  {journal} {\bibinfo
  {journal} {Nature}\ }\textbf {\bibinfo {volume} {551}},\ \bibinfo {pages}
  {579} (\bibinfo {year} {2017})}\BibitemShut {NoStop}%
\bibitem [{\citenamefont {Dom{\'i}nguez}\ \emph {et~al.}(2016)\citenamefont
  {Dom{\'i}nguez}, \citenamefont {Gonz{\'a}lez}, \citenamefont {Segnorile},\
  and\ \citenamefont {Zamar}}]{Dominguez.2016}%
  \BibitemOpen
  \bibfield  {author} {\bibinfo {author} {\bibfnamefont {F.~D.}\ \bibnamefont
  {Dom{\'i}nguez}}, \bibinfo {author} {\bibfnamefont {C.~E.}\ \bibnamefont
  {Gonz{\'a}lez}}, \bibinfo {author} {\bibfnamefont {H.~H.}\ \bibnamefont
  {Segnorile}},\ and\ \bibinfo {author} {\bibfnamefont {R.~C.}\ \bibnamefont
  {Zamar}},\ }\href {https://doi.org/10.1103/PhysRevA.93.022120} {\bibfield
  {journal} {\bibinfo  {journal} {Physical Review A}\ }\textbf {\bibinfo
  {volume} {93}},\ \bibinfo {pages} {022120} (\bibinfo {year}
  {2016})}\BibitemShut {NoStop}%
\bibitem [{\citenamefont {Mizushima}\ \emph {et~al.}(2017)\citenamefont
  {Mizushima}, \citenamefont {Goto},\ and\ \citenamefont
  {Sato}}]{Mizushima.2017}%
  \BibitemOpen
  \bibfield  {author} {\bibinfo {author} {\bibfnamefont {K.}~\bibnamefont
  {Mizushima}}, \bibinfo {author} {\bibfnamefont {H.}~\bibnamefont {Goto}},\
  and\ \bibinfo {author} {\bibfnamefont {R.}~\bibnamefont {Sato}},\ }\href
  {https://doi.org/10.1063/1.5007231} {\bibfield  {journal} {\bibinfo
  {journal} {Applied Physics Letters}\ }\textbf {\bibinfo {volume} {111}},\
  \bibinfo {pages} {172406} (\bibinfo {year} {2017})}\BibitemShut {NoStop}%
\bibitem [{\citenamefont {Borovik-Romanov}\ \emph {et~al.}(1984)\citenamefont
  {Borovik-Romanov}, \citenamefont {Bun'kov}, \citenamefont {Dumesh},
  \citenamefont {Kurkin}, \citenamefont {Petrov},\ and\ \citenamefont
  {Chekmarev}}]{BorovikRomanov.1984}%
  \BibitemOpen
  \bibfield  {author} {\bibinfo {author} {\bibfnamefont {A.~S.}\ \bibnamefont
  {Borovik-Romanov}}, \bibinfo {author} {\bibfnamefont {Y.~M.}\ \bibnamefont
  {Bun'kov}}, \bibinfo {author} {\bibfnamefont {B.~S.}\ \bibnamefont {Dumesh}},
  \bibinfo {author} {\bibfnamefont {M.~I.}\ \bibnamefont {Kurkin}}, \bibinfo
  {author} {\bibfnamefont {M.~P.}\ \bibnamefont {Petrov}},\ and\ \bibinfo
  {author} {\bibfnamefont {V.~P.}\ \bibnamefont {Chekmarev}},\ }\href
  {https://doi.org/10.1070/PU1984v027n04ABEH004041} {\bibfield  {journal}
  {\bibinfo  {journal} {Soviet Physics Uspekhi}\ }\textbf {\bibinfo {volume}
  {27}},\ \bibinfo {pages} {235} (\bibinfo {year} {1984})}\BibitemShut
  {NoStop}%
\bibitem [{\citenamefont {Papavassiliou}\ \emph {et~al.}(2000)\citenamefont
  {Papavassiliou}, \citenamefont {Fardis}, \citenamefont {Belesi},
  \citenamefont {Maris}, \citenamefont {Kallias}, \citenamefont {Pissas},
  \citenamefont {Niarchos}, \citenamefont {Dimitropoulos},\ and\ \citenamefont
  {Dolinsek}}]{Papavassiliou.2000}%
  \BibitemOpen
  \bibfield  {author} {\bibinfo {author} {\bibfnamefont {G.}~\bibnamefont
  {Papavassiliou}}, \bibinfo {author} {\bibfnamefont {M.}~\bibnamefont
  {Fardis}}, \bibinfo {author} {\bibfnamefont {M.}~\bibnamefont {Belesi}},
  \bibinfo {author} {\bibfnamefont {T.~G.}\ \bibnamefont {Maris}}, \bibinfo
  {author} {\bibfnamefont {G.}~\bibnamefont {Kallias}}, \bibinfo {author}
  {\bibfnamefont {M.}~\bibnamefont {Pissas}}, \bibinfo {author} {\bibfnamefont
  {D.}~\bibnamefont {Niarchos}}, \bibinfo {author} {\bibfnamefont
  {C.}~\bibnamefont {Dimitropoulos}},\ and\ \bibinfo {author} {\bibfnamefont
  {J.}~\bibnamefont {Dolinsek}},\ }\href
  {https://doi.org/10.1103/PhysRevLett.84.761} {\bibfield  {journal} {\bibinfo
  {journal} {Physical review letters}\ }\textbf {\bibinfo {volume} {84}},\
  \bibinfo {pages} {761} (\bibinfo {year} {2000})}\BibitemShut {NoStop}%
\bibitem [{\citenamefont {Panopoulos}\ \emph {et~al.}(2018)\citenamefont
  {Panopoulos}, \citenamefont {Pissas}, \citenamefont {Kim}, \citenamefont
  {Kim}, \citenamefont {Yoo}, \citenamefont {Hassan}, \citenamefont {AlWahedi},
  \citenamefont {Alhassan}, \citenamefont {Fardis}, \citenamefont {Boukos},\
  and\ \citenamefont {Papavassiliou}}]{Panopoulos.2018}%
  \BibitemOpen
  \bibfield  {author} {\bibinfo {author} {\bibfnamefont {N.}~\bibnamefont
  {Panopoulos}}, \bibinfo {author} {\bibfnamefont {M.}~\bibnamefont {Pissas}},
  \bibinfo {author} {\bibfnamefont {H.~J.}\ \bibnamefont {Kim}}, \bibinfo
  {author} {\bibfnamefont {J.-G.}\ \bibnamefont {Kim}}, \bibinfo {author}
  {\bibfnamefont {S.~J.}\ \bibnamefont {Yoo}}, \bibinfo {author} {\bibfnamefont
  {J.}~\bibnamefont {Hassan}}, \bibinfo {author} {\bibfnamefont
  {Y.}~\bibnamefont {AlWahedi}}, \bibinfo {author} {\bibfnamefont
  {S.}~\bibnamefont {Alhassan}}, \bibinfo {author} {\bibfnamefont
  {M.}~\bibnamefont {Fardis}}, \bibinfo {author} {\bibfnamefont
  {N.}~\bibnamefont {Boukos}},\ and\ \bibinfo {author} {\bibfnamefont
  {G.}~\bibnamefont {Papavassiliou}},\ }\href
  {https://doi.org/10.1038/s41535-018-0093-4} {\bibfield  {journal} {\bibinfo
  {journal} {npj Quantum Materials}\ }\textbf {\bibinfo {volume} {3}},\
  \bibinfo {pages} {57007} (\bibinfo {year} {2018})}\BibitemShut {NoStop}%
\bibitem [{\citenamefont {Germov}\ \emph {et~al.}(2019)\citenamefont {Germov},
  \citenamefont {Mikhalev}, \citenamefont {Volkova}, \citenamefont
  {Gerashchenko}, \citenamefont {Konstantinova},\ and\ \citenamefont
  {Leonidov}}]{Germov.2019}%
  \BibitemOpen
  \bibfield  {author} {\bibinfo {author} {\bibfnamefont {A.~Y.}\ \bibnamefont
  {Germov}}, \bibinfo {author} {\bibfnamefont {K.~N.}\ \bibnamefont
  {Mikhalev}}, \bibinfo {author} {\bibfnamefont {Z.~N.}\ \bibnamefont
  {Volkova}}, \bibinfo {author} {\bibfnamefont {A.~P.}\ \bibnamefont
  {Gerashchenko}}, \bibinfo {author} {\bibfnamefont {E.~I.}\ \bibnamefont
  {Konstantinova}},\ and\ \bibinfo {author} {\bibfnamefont {I.~A.}\
  \bibnamefont {Leonidov}},\ }\href {https://doi.org/10.1134/S002136401904009X}
  {\bibfield  {journal} {\bibinfo  {journal} {JETP Letters}\ }\textbf {\bibinfo
  {volume} {109}},\ \bibinfo {pages} {252} (\bibinfo {year}
  {2019})}\BibitemShut {NoStop}%
\bibitem [{\citenamefont {Tomka}\ \emph {et~al.}(1998)\citenamefont {Tomka},
  \citenamefont {Riedi}, \citenamefont {Kapusta}, \citenamefont {Balakrishnan},
  \citenamefont {Paul}, \citenamefont {Lees},\ and\ \citenamefont
  {Barratt}}]{Tomka.1998}%
  \BibitemOpen
  \bibfield  {author} {\bibinfo {author} {\bibfnamefont {G.~J.}\ \bibnamefont
  {Tomka}}, \bibinfo {author} {\bibfnamefont {P.~C.}\ \bibnamefont {Riedi}},
  \bibinfo {author} {\bibfnamefont {C.}~\bibnamefont {Kapusta}}, \bibinfo
  {author} {\bibfnamefont {G.}~\bibnamefont {Balakrishnan}}, \bibinfo {author}
  {\bibfnamefont {D.~M.}\ \bibnamefont {Paul}}, \bibinfo {author}
  {\bibfnamefont {M.~R.}\ \bibnamefont {Lees}},\ and\ \bibinfo {author}
  {\bibfnamefont {J.}~\bibnamefont {Barratt}},\ }\href
  {https://doi.org/10.1063/1.367626} {\bibfield  {journal} {\bibinfo  {journal}
  {Journal of Applied Physics}\ }\textbf {\bibinfo {volume} {83}},\ \bibinfo
  {pages} {7151} (\bibinfo {year} {1998})}\BibitemShut {NoStop}%
\bibitem [{\citenamefont {Mazur}(2012)}]{Mazur.2012}%
  \BibitemOpen
  \bibfield  {author} {\bibinfo {author} {\bibfnamefont {A.~S.}\ \bibnamefont
  {Mazur}},\ }\href {https://doi.org/10.1134/S1063783412110194} {\bibfield
  {journal} {\bibinfo  {journal} {Physics of the Solid State}\ }\textbf
  {\bibinfo {volume} {54}},\ \bibinfo {pages} {2222} (\bibinfo {year}
  {2012})}\BibitemShut {NoStop}%
\bibitem [{\citenamefont {Savosta}\ and\ \citenamefont
  {Nov{\'a}k}(2001)}]{Savosta.2001}%
  \BibitemOpen
  \bibfield  {author} {\bibinfo {author} {\bibfnamefont {M.~M.}\ \bibnamefont
  {Savosta}}\ and\ \bibinfo {author} {\bibfnamefont {P.}~\bibnamefont
  {Nov{\'a}k}},\ }\href {https://doi.org/10.1103/PhysRevLett.87.137204}
  {\bibfield  {journal} {\bibinfo  {journal} {Physical review letters}\
  }\textbf {\bibinfo {volume} {87}},\ \bibinfo {pages} {137204} (\bibinfo
  {year} {2001})}\BibitemShut {NoStop}%
\bibitem [{\citenamefont {Zviadadze}\ \emph {et~al.}(2013)\citenamefont
  {Zviadadze}, \citenamefont {Mamniashvili}, \citenamefont {Gegechkori},
  \citenamefont {Akhalkatsi},\ and\ \citenamefont
  {Gavasheli}}]{Zviadadze.2013}%
  \BibitemOpen
  \bibfield  {author} {\bibinfo {author} {\bibfnamefont {M.~D.}\ \bibnamefont
  {Zviadadze}}, \bibinfo {author} {\bibfnamefont {G.~I.}\ \bibnamefont
  {Mamniashvili}}, \bibinfo {author} {\bibfnamefont {T.~O.}\ \bibnamefont
  {Gegechkori}}, \bibinfo {author} {\bibfnamefont {A.~M.}\ \bibnamefont
  {Akhalkatsi}},\ and\ \bibinfo {author} {\bibfnamefont {T.~A.}\ \bibnamefont
  {Gavasheli}},\ }\href {https://doi.org/10.1007/s10948-012-2039-6} {\bibfield
  {journal} {\bibinfo  {journal} {J Supercond Nov Magn}\ }\textbf {\bibinfo {volume} {26}},\ \bibinfo {pages} {1405}
  (\bibinfo {year} {2013})}\BibitemShut {NoStop}%
\bibitem [{\citenamefont {Kuz'min}\ and\ \citenamefont
  {Kolesenko}(2001)}]{Kuzmin.2001}%
  \BibitemOpen
  \bibfield  {author} {\bibinfo {author} {\bibfnamefont {V.~S.}\ \bibnamefont
  {Kuz'min}}\ and\ \bibinfo {author} {\bibfnamefont {V.~M.}\ \bibnamefont
  {Kolesenko}},\ }\href {https://doi.org/10.1023/A:1011966826581} {\bibfield
  {journal} {\bibinfo  {journal} {J. Appl. Spectrosc.}\ }\textbf
  {\bibinfo {volume} {68}},\ \bibinfo {pages} {480} (\bibinfo {year}
  {2001})}\BibitemShut {NoStop}%
\bibitem [{\citenamefont {Kaiser}(1981)}]{Kaiser.1981}%
  \BibitemOpen
  \bibfield  {author} {\bibinfo {author} {\bibfnamefont {R.}~\bibnamefont
  {Kaiser}},\ }\href {https://doi.org/10.1016/0022-2364(81)90013-5} {\bibfield
  {journal} {\bibinfo  {journal} {Journal of magnetic resonance}\ }\textbf
  {\bibinfo {volume} {42}},\ \bibinfo {pages} {103} (\bibinfo {year}
  {1981})}\BibitemShut {NoStop}%
\bibitem [{\citenamefont {Kuz'min}\ and\ \citenamefont
  {Kolesenko}(2012)}]{Kuzmin.2012}%
  \BibitemOpen
  \bibfield  {author} {\bibinfo {author} {\bibfnamefont {V.~S.}\ \bibnamefont
  {Kuz'min}}\ and\ \bibinfo {author} {\bibfnamefont {V.~M.}\ \bibnamefont
  {Kolesenko}},\ }\href {https://doi.org/10.1007/s10812-012-9613-3} {\bibfield
  {journal} {\bibinfo  {journal} {J. Appl. Spectrosc.}\ }\textbf
  {\bibinfo {volume} {79}},\ \bibinfo {pages} {390} (\bibinfo {year}
  {2012})}\BibitemShut {NoStop}%
\bibitem [{\citenamefont {Kuz'min}\ and\ \citenamefont
  {Kolesenko}(2005)}]{Kuzmin.2005}%
  \BibitemOpen
  \bibfield  {author} {\bibinfo {author} {\bibfnamefont {V.~S.}\ \bibnamefont
  {Kuz'min}}\ and\ \bibinfo {author} {\bibfnamefont {V.~M.}\ \bibnamefont
  {Kolesenko}},\ }\href {https://doi.org/10.1134/1.2131148} {\bibfield
  {journal} {\bibinfo  {journal} {Physics of the Solid State}\ }\textbf
  {\bibinfo {volume} {47}},\ \bibinfo {pages} {2077} (\bibinfo {year}
  {2005})}\BibitemShut {NoStop}%
\bibitem [{\citenamefont {Pincus}\ \emph {et~al.}(1968)\citenamefont {Pincus},
  \citenamefont {Jaccarino}, \citenamefont {Hone},\ and\ \citenamefont
  {Ngwe}}]{Pincus.1968}%
  \BibitemOpen
  \bibfield  {author} {\bibinfo {author} {\bibfnamefont {P.}~\bibnamefont
  {Pincus}}, \bibinfo {author} {\bibfnamefont {V.}~\bibnamefont {Jaccarino}},
  \bibinfo {author} {\bibfnamefont {D.}~\bibnamefont {Hone}},\ and\ \bibinfo
  {author} {\bibfnamefont {T.}~\bibnamefont {Ngwe}},\ }\href
  {https://doi.org/10.1016/0375-9601(68)91334-0} {\bibfield  {journal}
  {\bibinfo  {journal} {Physics Letters A}\ }\textbf {\bibinfo {volume} {27}},\
  \bibinfo {pages} {54} (\bibinfo {year} {1968})}\BibitemShut {NoStop}%
\bibitem [{\citenamefont {Tagirov}\ \emph {et~al.}(2014)\citenamefont
  {Tagirov}, \citenamefont {Alakshin}, \citenamefont {Bunkov}, \citenamefont
  {Gazizulin}, \citenamefont {Gazizulina}, \citenamefont {Isaenko},
  \citenamefont {Klochkov}, \citenamefont {Safin}, \citenamefont {Safiullin},\
  and\ \citenamefont {Zhurkov}}]{Tagirov.2014}%
  \BibitemOpen
  \bibfield  {author} {\bibinfo {author} {\bibfnamefont {M.~S.}\ \bibnamefont
  {Tagirov}}, \bibinfo {author} {\bibfnamefont {E.~M.}\ \bibnamefont
  {Alakshin}}, \bibinfo {author} {\bibfnamefont {Y.~M.}\ \bibnamefont
  {Bunkov}}, \bibinfo {author} {\bibfnamefont {R.~R.}\ \bibnamefont
  {Gazizulin}}, \bibinfo {author} {\bibfnamefont {A.~M.}\ \bibnamefont
  {Gazizulina}}, \bibinfo {author} {\bibfnamefont {L.~I.}\ \bibnamefont
  {Isaenko}}, \bibinfo {author} {\bibfnamefont {A.~V.}\ \bibnamefont
  {Klochkov}}, \bibinfo {author} {\bibfnamefont {T.~R.}\ \bibnamefont {Safin}},
  \bibinfo {author} {\bibfnamefont {K.~R.}\ \bibnamefont {Safiullin}},\ and\
  \bibinfo {author} {\bibfnamefont {S.~A.}\ \bibnamefont {Zhurkov}},\ }\href
  {https://doi.org/10.1007/s10909-013-1085-1} {\bibfield  {journal} {\bibinfo
  {journal} {Journal of Low Temperature Physics}\ }\textbf {\bibinfo {volume}
  {175}},\ \bibinfo {pages} {167} (\bibinfo {year} {2014})}\BibitemShut
  {NoStop}%
\bibitem [{\citenamefont {Bun'kov}\ \emph {et~al.}(1974)\citenamefont
  {Bun'kov}, \citenamefont {Dumesh},\ and\ \citenamefont
  {Kurkin}}]{Bunkov.1974}%
  \BibitemOpen
  \bibfield  {author} {\bibinfo {author} {\bibfnamefont {Y.}~\bibnamefont
  {Bun'kov}}, \bibinfo {author} {\bibfnamefont {B.~S.}\ \bibnamefont
  {Dumesh}},\ and\ \bibinfo {author} {\bibfnamefont {M.~I.}\ \bibnamefont
  {Kurkin}},\ }\href@noop {} {\bibfield  {journal} {\bibinfo  {journal} {JETP
  Letters}\ }\textbf {\bibinfo {volume} {19}},\ \bibinfo {pages} {132}
  (\bibinfo {year} {1974})}\BibitemShut {NoStop}%
\bibitem [{\citenamefont {Zviadadze}\ \emph {et~al.}(2015)\citenamefont
  {Zviadadze}, \citenamefont {Mamniashvili}, \citenamefont {Akhalkatsi},\ and\
  \citenamefont {Menabde}}]{Zviadadze.2015}%
  \BibitemOpen
  \bibfield  {author} {\bibinfo {author} {\bibfnamefont {M.}~\bibnamefont
  {Zviadadze}}, \bibinfo {author} {\bibfnamefont {G.}~\bibnamefont
  {Mamniashvili}}, \bibinfo {author} {\bibfnamefont {A.}~\bibnamefont
  {Akhalkatsi}},\ and\ \bibinfo {author} {\bibfnamefont {M.}~\bibnamefont
  {Menabde}},\ }\href {https://doi.org/10.1007/s10948-014-2757-z} {\bibfield
  {journal} {\bibinfo  {journal} {J Supercond Nov Magn}\ }\textbf {\bibinfo {volume} {28}},\ \bibinfo {pages} {927}
  (\bibinfo {year} {2015})}\BibitemShut {NoStop}%
\bibitem [{\citenamefont {Akhalkatsi}\ \emph {et~al.}(2002)\citenamefont
  {Akhalkatsi}, \citenamefont {Mamniashvili}, \citenamefont {Gegechkori},\ and\
  \citenamefont {{Ben Ezra}}}]{Akhalkatsi.2002}%
  \BibitemOpen
  \bibfield  {author} {\bibinfo {author} {\bibfnamefont {A.~M.}\ \bibnamefont
  {Akhalkatsi}}, \bibinfo {author} {\bibfnamefont {G.}~\bibnamefont
  {Mamniashvili}}, \bibinfo {author} {\bibfnamefont {T.~O.}\ \bibnamefont
  {Gegechkori}},\ and\ \bibinfo {author} {\bibfnamefont {S.}~\bibnamefont {{Ben
  Ezra}}},\ }\href@noop {} {\bibfield  {journal} {\bibinfo  {journal} {The
  Physics of Metals and Metallography}\ }\textbf {\bibinfo {volume} {94}},\
  \bibinfo {pages} {33} (\bibinfo {year} {2002})}\BibitemShut {NoStop}%
\bibitem [{\citenamefont {Mamniashvili}\ \emph {et~al.}(2015)\citenamefont
  {Mamniashvili}, \citenamefont {Gegechkori}, \citenamefont {Akhalkatsi},\ and\
  \citenamefont {Gavasheli}}]{Mamniashvili.2015}%
  \BibitemOpen
  \bibfield  {author} {\bibinfo {author} {\bibfnamefont {G.}~\bibnamefont
  {Mamniashvili}}, \bibinfo {author} {\bibfnamefont {T.}~\bibnamefont
  {Gegechkori}}, \bibinfo {author} {\bibfnamefont {A.}~\bibnamefont
  {Akhalkatsi}},\ and\ \bibinfo {author} {\bibfnamefont {T.}~\bibnamefont
  {Gavasheli}},\ }\href {https://doi.org/10.1007/s10948-014-2812-9} {\bibfield
  {journal} {\bibinfo  {journal} {J Supercond Nov Magn}\ }\textbf {\bibinfo {volume} {28}},\ \bibinfo {pages} {911}
  (\bibinfo {year} {2015})}\BibitemShut {NoStop}%
\bibitem [{\citenamefont {Shakhmuratova}\ \emph {et~al.}(1997)\citenamefont
  {Shakhmuratova}, \citenamefont {Fowler},\ and\ \citenamefont
  {Chaplin}}]{Shakhmuratova.1997}%
  \BibitemOpen
  \bibfield  {author} {\bibinfo {author} {\bibfnamefont {L.~N.}\ \bibnamefont
  {Shakhmuratova}}, \bibinfo {author} {\bibfnamefont {D.~K.}\ \bibnamefont
  {Fowler}},\ and\ \bibinfo {author} {\bibfnamefont {D.~H.}\ \bibnamefont
  {Chaplin}},\ }\href {https://doi.org/10.1103/PhysRevA.55.2955} {\bibfield
  {journal} {\bibinfo  {journal} {Physical Review A}\ }\textbf {\bibinfo
  {volume} {55}},\ \bibinfo {pages} {2955} (\bibinfo {year}
  {1997})}\BibitemShut {NoStop}%
\bibitem [{\citenamefont {Kuz'min}\ and\ \citenamefont
  {Kolesenko}(2006)}]{Kuzmin.2006}%
  \BibitemOpen
  \bibfield  {author} {\bibinfo {author} {\bibfnamefont {V.~S.}\ \bibnamefont
  {Kuz'min}}\ and\ \bibinfo {author} {\bibfnamefont {V.~M.}\ \bibnamefont
  {Kolesenko}},\ }\href {https://doi.org/10.1007/s10812-006-0080-6} {\bibfield
  {journal} {\bibinfo  {journal} {J. Appl. Spectrosc.}\ }\textbf
  {\bibinfo {volume} {73}},\ \bibinfo {pages} {340} (\bibinfo {year}
  {2006})}\BibitemShut {NoStop}%
\bibitem [{\citenamefont {Fedoruk}(2002)}]{Fedoruk.2002}%
  \BibitemOpen
  \bibfield  {author} {\bibinfo {author} {\bibfnamefont {G.~G.}\ \bibnamefont
  {Fedoruk}},\ }\href {https://doi.org/10.1023/A:1016113814743} {\bibfield
  {journal} {\bibinfo  {journal} {J. Appl. Spectrosc.}\ }\textbf
  {\bibinfo {volume} {69}},\ \bibinfo {pages} {161} (\bibinfo {year}
  {2002})}\BibitemShut {NoStop}%
\bibitem [{\citenamefont {Saiko}\ \emph
  {et~al.}(2018{\natexlab{a}})\citenamefont {Saiko}, \citenamefont
  {Markevich},\ and\ \citenamefont {Fedaruk}}]{Saiko.2018}%
  \BibitemOpen
  \bibfield  {author} {\bibinfo {author} {\bibfnamefont {A.~P.}\ \bibnamefont
  {Saiko}}, \bibinfo {author} {\bibfnamefont {S.~A.}\ \bibnamefont
  {Markevich}},\ and\ \bibinfo {author} {\bibfnamefont {R.}~\bibnamefont
  {Fedaruk}},\ }\href {https://doi.org/10.1103/PhysRevA.98.043814} {\bibfield
  {journal} {\bibinfo  {journal} {Physical Review A}\ }\textbf {\bibinfo
  {volume} {98}},\ \bibinfo {pages} {043814} (\bibinfo {year}
  {2018}{\natexlab{a}})}\BibitemShut {NoStop}%
\bibitem [{\citenamefont {Saiko}\ \emph
  {et~al.}(2018{\natexlab{b}})\citenamefont {Saiko}, \citenamefont {Fedaruk},\
  and\ \citenamefont {Markevich}}]{Saiko.2018b}%
  \BibitemOpen
  \bibfield  {author} {\bibinfo {author} {\bibfnamefont {A.~P.}\ \bibnamefont
  {Saiko}}, \bibinfo {author} {\bibfnamefont {R.}~\bibnamefont {Fedaruk}},\
  and\ \bibinfo {author} {\bibfnamefont {S.~A.}\ \bibnamefont {Markevich}},\
  }\href {https://doi.org/10.1016/j.jmr.2018.02.003} {\bibfield  {journal}
  {\bibinfo  {journal} {Journal of magnetic resonance}\ }\textbf {\bibinfo
  {volume} {290}},\ \bibinfo {pages} {60} (\bibinfo {year}
  {2018}{\natexlab{b}})}\BibitemShut {NoStop}%
\bibitem [{\citenamefont {Khasanov}\ \emph {et~al.}(2007)\citenamefont
  {Khasanov}, \citenamefont {Fedotova},\ and\ \citenamefont
  {Samartsev}}]{Khasanov.2007}%
  \BibitemOpen
  \bibfield  {author} {\bibinfo {author} {\bibfnamefont {O.~K.}\ \bibnamefont
  {Khasanov}}, \bibinfo {author} {\bibfnamefont {O.~M.}\ \bibnamefont
  {Fedotova}},\ and\ \bibinfo {author} {\bibfnamefont {V.~V.}\ \bibnamefont
  {Samartsev}},\ }\href {https://doi.org/10.1016/j.jlumin.2007.02.049}
  {\bibfield  {journal} {\bibinfo  {journal} {Journal of Luminescence}\
  }\textbf {\bibinfo {volume} {127}},\ \bibinfo {pages} {55} (\bibinfo {year}
  {2007})}\BibitemShut {NoStop}%
\bibitem [{\citenamefont {Khasanov}\ \emph {et~al.}(2003)\citenamefont
  {Khasanov}, \citenamefont {Rusetsky}, \citenamefont {Samartsev},
  \citenamefont {Smirnova},\ and\ \citenamefont {{Fedotova
  O.M.}}}]{Khasanov.2003}%
  \BibitemOpen
  \bibfield  {author} {\bibinfo {author} {\bibfnamefont {O.}~\bibnamefont
  {Khasanov}}, \bibinfo {author} {\bibfnamefont {G.~A.}\ \bibnamefont
  {Rusetsky}}, \bibinfo {author} {\bibfnamefont {V.~V.}\ \bibnamefont
  {Samartsev}}, \bibinfo {author} {\bibfnamefont {T.~V.}\ \bibnamefont
  {Smirnova}},\ and\ \bibinfo {author} {\bibnamefont {{Fedotova O.M.}}},\
  }\href@noop {} {\bibfield  {journal} {\bibinfo  {journal} {Proceedings of
  SPIE - The International Society for Optical Engineering}\ }\textbf {\bibinfo
  {volume} {5402}},\ \bibinfo {pages} {347} (\bibinfo {year}
  {2004})}\BibitemShut {NoStop}%
\bibitem [{\citenamefont {Schmitt}\ and\ \citenamefont
  {Kehrein}(2016)}]{Schmitt.2016}%
  \BibitemOpen
  \bibfield  {author} {\bibinfo {author} {\bibfnamefont {M.}~\bibnamefont
  {Schmitt}}\ and\ \bibinfo {author} {\bibfnamefont {S.}~\bibnamefont
  {Kehrein}},\ }\href {https://doi.org/10.1209/0295-5075/115/50001} {\bibfield
  {journal} {\bibinfo  {journal} {EPL (Europhysics Letters)}\ }\textbf
  {\bibinfo {volume} {115}},\ \bibinfo {pages} {50001} (\bibinfo {year}
  {2016})}\BibitemShut {NoStop}%
\bibitem [{\citenamefont {Schmitt}\ and\ \citenamefont
  {Kehrein}(2018)}]{Schmitt.2018}%
  \BibitemOpen
  \bibfield  {author} {\bibinfo {author} {\bibfnamefont {M.}~\bibnamefont
  {Schmitt}}\ and\ \bibinfo {author} {\bibfnamefont {S.}~\bibnamefont
  {Kehrein}},\ }\href {https://doi.org/10.1103/PhysRevB.98.180301} {\bibfield
  {journal} {\bibinfo  {journal} {Phys. Rev. B}\ }\textbf {\bibinfo {volume}
  {98}},\ \bibinfo {pages} {180301(R)} (\bibinfo {year} {2018})}\BibitemShut
  {NoStop}%
\bibitem{ernst1990principles} R. Ernst, G. Bodenhausen, and A. Wokaun, \emph{Principles of
Nuclear Magnetic Resonance in One and Two Dimensions},
International Series of Monographs on Chemistry (Clarendon
Press, 1991)%
\bibitem [{\citenamefont {Khasanov}\ \emph {et~al.}(2019)\citenamefont
  {Khasanov}, \citenamefont {Fedotova}, \citenamefont {Rusetsky},\ and\
  \citenamefont {Nikiforov}}]{Khasanov.2019}%
  \BibitemOpen
  \bibfield  {author} {\bibinfo {author} {\bibfnamefont {O.}~\bibnamefont
  {Khasanov}}, \bibinfo {author} {\bibfnamefont {O.}~\bibnamefont {Fedotova}},
  \bibinfo {author} {\bibfnamefont {G.}~\bibnamefont {Rusetsky}},\ and\
  \bibinfo {author} {\bibfnamefont {V.}~\bibnamefont {Nikiforov}},\ }\href
  {https://doi.org/10.1088/1555-6611/ab4bd2} {\bibfield  {journal} {\bibinfo
  {journal} {Laser Physics}\ }\textbf {\bibinfo {volume} {29}},\ \bibinfo
  {pages} {124011} (\bibinfo {year} {2019})}\BibitemShut {NoStop}%
\bibitem [{\citenamefont {Pokazan'ev}\ and\ \citenamefont
  {Yakub}(1977)}]{Pokazanev.1977}%
  \BibitemOpen
  \bibfield  {author} {\bibinfo {author} {\bibfnamefont {V.~G.}\ \bibnamefont
  {Pokazan'ev}}\ and\ \bibinfo {author} {\bibfnamefont {L.~I.}\ \bibnamefont
  {Yakub}},\ }\href@noop {} {\bibfield  {journal} {\bibinfo  {journal} {Zh. Eksp. Teor. Fiz.}\ }\textbf {\bibinfo {volume} {73}},\
  \bibinfo {pages} {221} (\bibinfo {year} {1977})}[Sov. Phys. JETP 46, 114 (1977)]\BibitemShut {NoStop}%
\bibitem [{\citenamefont {Kiliptari}\ and\ \citenamefont
  {Tsifrinovich}(1998)}]{Kiliptari.1998}%
  \BibitemOpen
  \bibfield  {author} {\bibinfo {author} {\bibfnamefont {I.~G.}\ \bibnamefont
  {Kiliptari}}\ and\ \bibinfo {author} {\bibfnamefont {V.~I.}\ \bibnamefont
  {Tsifrinovich}},\ }\href {https://doi.org/10.1103/PhysRevB.57.11554}
  {\bibfield  {journal} {\bibinfo  {journal} {Physical Review B}\ }\textbf
  {\bibinfo {volume} {57}},\ \bibinfo {pages} {11554} (\bibinfo {year}
  {1998})}\BibitemShut {NoStop}%
\bibitem [{\citenamefont {Savosta}\ \emph {et~al.}(2003)\citenamefont
  {Savosta}, \citenamefont {Doroshev}, \citenamefont {Kamenev}, \citenamefont
  {Borodin}, \citenamefont {Tarasenko}, \citenamefont {Mazur},\ and\
  \citenamefont {Mary{\v{s}}ko}}]{Savosta.2003}%
  \BibitemOpen
  \bibfield  {author} {\bibinfo {author} {\bibfnamefont {M.~M.}\ \bibnamefont
  {Savosta}}, \bibinfo {author} {\bibfnamefont {V.~D.}\ \bibnamefont
  {Doroshev}}, \bibinfo {author} {\bibfnamefont {V.~I.}\ \bibnamefont
  {Kamenev}}, \bibinfo {author} {\bibfnamefont {V.~A.}\ \bibnamefont
  {Borodin}}, \bibinfo {author} {\bibfnamefont {T.~N.}\ \bibnamefont
  {Tarasenko}}, \bibinfo {author} {\bibfnamefont {A.~S.}\ \bibnamefont
  {Mazur}},\ and\ \bibinfo {author} {\bibfnamefont {M.}~\bibnamefont
  {Mary{\v{s}}ko}},\ }\href {https://doi.org/10.1134/1.1618342} {\bibfield
  {journal} {\bibinfo  {journal} {J. Exp. Theor.
Phys.}\ }\textbf {\bibinfo {volume} {97}},\ \bibinfo {pages} {573}
  (\bibinfo {year} {2003})}\BibitemShut {NoStop}%
\bibitem [{\citenamefont {Kuz'min}\ and\ \citenamefont
  {Fedoruk}(2001)}]{Kuzmin.2001b}%
  \BibitemOpen
  \bibfield  {author} {\bibinfo {author} {\bibfnamefont {V.~S.}\ \bibnamefont
  {Kuz'min}}\ and\ \bibinfo {author} {\bibfnamefont {G.~G.}\ \bibnamefont
  {Fedoruk}},\ }\href@noop {} {\emph {\bibinfo {title} {Nonstationary coherent
  phenomena in paramagnetic spin systems.}}}\ (\bibinfo  {publisher} {BSU},\
  \bibinfo {address} {Minsk},\ \bibinfo {year} {2001})\BibitemShut {NoStop}%
\bibitem [{\citenamefont {Nayfeh}(2011)}]{Nayfeh.2011}%
  \BibitemOpen
  \bibfield  {author} {\bibinfo {author} {\bibfnamefont {A.~H.}\ \bibnamefont
  {Nayfeh}},\ }\href@noop {} {\emph {\bibinfo {title} {Introduction to
  Perturbation Techniques}}}\ (\bibinfo  {publisher} {{John Wiley {\&} Sons}},\
  \bibinfo {address} {Hoboken},\ \bibinfo {year} {2011})\BibitemShut {NoStop}%
\bibitem [{\citenamefont {Kuz'min}\ \emph {et~al.}(1990)\citenamefont
  {Kuz'min}, \citenamefont {Rutkovskii}, \citenamefont {Saiko}, \citenamefont
  {Tarasevich},\ and\ \citenamefont {Fedoruk}}]{Kuzmin.1990}%
  V. S. Kuz’min, I. Z. Rutkovskii, A. P. Saiko, A. D. Tarasevich, and G. G. Fedoruk, Zh. Eksp. Teor. Fiz. 97, 880 (1990) [Sov. Phys. JETP 70, 493 (1990)]
%
\bibitem [{\citenamefont {Berzhanskii}\ \emph {et~al.}(2020)\citenamefont
  {Berzhanskii}, \citenamefont {Gippius}, \citenamefont {Gorbovanov},
  \citenamefont {Zhurenko},\ and\ \citenamefont
  {Polulyakh}}]{Berzhanskii.2020}%
  \BibitemOpen
  \bibfield  {author} {\bibinfo {author} {\bibfnamefont {V.~N.}\ \bibnamefont
  {Berzhanskii}}, \bibinfo {author} {\bibfnamefont {A.~A.}\ \bibnamefont
  {Gippius}}, \bibinfo {author} {\bibfnamefont {A.~I.}\ \bibnamefont
  {Gorbovanov}}, \bibinfo {author} {\bibfnamefont {S.~V.}\ \bibnamefont
  {Zhurenko}},\ and\ \bibinfo {author} {\bibfnamefont {S.~N.}\ \bibnamefont
  {Polulyakh}},\ }\href {https://doi.org/10.1134/s106377611912015x} {\bibfield
  {journal} {\bibinfo  {journal} {J. Exp. Theor. Phys.}\ }\textbf {\bibinfo {volume} {130}},\ \bibinfo {pages} {101}
  (\bibinfo {year} {2020})}\BibitemShut {NoStop}%
\bibitem [{\citenamefont {Kuz'min}\ and\ \citenamefont
  {Kolesenko}(2002)}]{Kuzmin.2002}%
  \BibitemOpen
  \bibfield  {author} {\bibinfo {author} {\bibfnamefont {V.~S.}\ \bibnamefont
  {Kuz'min}}\ and\ \bibinfo {author} {\bibfnamefont {V.~M.}\ \bibnamefont
  {Kolesenko}},\ }\href@noop {} {\bibfield  {journal} {\bibinfo  {journal}
  {J. Appl. Spectrosc.}\ }\textbf {\bibinfo {volume} {69}},\
  \bibinfo {pages} {683} (\bibinfo {year} {2002})}\BibitemShut {NoStop}%
\bibitem [{\citenamefont {Kuzmin}\ and\ \citenamefont
  {Saiko}(1989)}]{Saiko.1989}%
  \BibitemOpen
  \bibfield  {author} {\bibinfo {author} {\bibfnamefont {V.~S.}\ \bibnamefont
  {Kuzmin}}\ and\ \bibinfo {author} {\bibfnamefont {A.~P.}\ \bibnamefont
  {Saiko}},\ }\href@noop {} {\bibfield  {journal} {\bibinfo  {journal} {Fizika
  Tverdogo Tela}\ }\textbf {\bibinfo {volume} {31}},\ \bibinfo {pages} {266}
  (\bibinfo {year} {1989})}\BibitemShut {NoStop}%
\bibitem [{\citenamefont {Kuzmin}\ \emph {et~al.}(1990)\citenamefont {Kuzmin},
  \citenamefont {Saiko},\ and\ \citenamefont {Fedoruk}}]{Saiko.1990}%
  \BibitemOpen
  \bibfield  {author} {\bibinfo {author} {\bibfnamefont {V.~S.}\ \bibnamefont
  {Kuzmin}}, \bibinfo {author} {\bibfnamefont {A.~P.}\ \bibnamefont {Saiko}},\
  and\ \bibinfo {author} {\bibfnamefont {G.~G.}\ \bibnamefont {Fedoruk}},\
  }\href@noop {} {\bibfield  {journal} {\bibinfo  {journal} {Fizika Tverdogo
  Tela}\ }\textbf {\bibinfo {volume} {32}},\ \bibinfo {pages} {608} (\bibinfo
  {year} {1990})}\BibitemShut {NoStop}%
\end{thebibliography}%

\end{document}